\documentclass[a4paper,aps,reprint,pra, 11pt, onecolumn, superscriptaddress,nofootinbib, showpacs, notitlepage]{revtex4-1} %\ifx\pdftexversion\undefined
\usepackage{times}
\usepackage{array}
\usepackage{amsmath}
\usepackage{amsfonts}
\usepackage{amssymb}
\usepackage{amsthm}
\usepackage{amsbsy}
\usepackage{stmaryrd}
\usepackage{pifont}
\usepackage{latexsym}
\usepackage{mathtools}
\usepackage{bbold}
\usepackage{gensymb}
\usepackage[all]{xy}
\usepackage{graphicx}
\usepackage{braket}

\usepackage{epstopdf}
\usepackage{epsfig}
\usepackage[labelformat=default, labelsep=default, justification=centerlast,  font=footnotesize]{caption}
\usepackage{verbatim}
\usepackage{multirow}
\usepackage{rotating}
\usepackage{minibox}
\usepackage{tabularx}

\usepackage[titletoc,toc,title]{appendix}
\usepackage{enumerate}
\usepackage{enumitem}
\usepackage{color}
\usepackage[dvipsnames]{xcolor}
\usepackage[right=2.75cm, left=2.75cm, top=3cm, bottom=3cm]{geometry}

\usepackage{comment}
\usepackage[colorlinks=True, linkcolor=Red,citecolor=Green]{hyperref}
\usepackage{sidecap}
\usepackage{bbm}
\usepackage{bm}

\DeclareCaptionJustification{justified}{\justifying}
\begin{document}

\title{Einstein's Equivalence principle for superpositions of gravitational fields and quantum reference frames}

\author{Flaminia Giacomini}%
 \email{fgiacomini@phys.ethz.ch}
\affiliation{%
Perimeter Institute for Theoretical Physics, 31 Caroline St. N, Waterloo, Ontario, N2L 2Y5, Canada
}%
\affiliation{Institute for Theoretical Physics, ETH Z{\"u}rich, Wolfgang-Pauli-Str. 27, Z{\"u}rich, Switzerland}

\author{\v{C}aslav Brukner}
\affiliation{%
	Vienna Center for Quantum Science and Technology (VCQ), Faculty of Physics, University of Vienna, Boltzmanngasse 5, A-1090 Vienna, Austria
}%
\affiliation{%
	Institute of Quantum Optics and Quantum Information (IQOQI), Austrian Academy of Sciences, Boltzmanngasse 3, A-1090 Vienna, Austria
}%

\begin{abstract}
	The Einstein Equivalence Principle (EEP), stating that all laws of physics take their special-relativistic form in any local inertial (classical) reference frame, lies at the core of general relativity. Because of its fundamental status, this principle could be a very powerful guide in formulating physical laws at regimes where both gravitational and quantum effects are relevant. The formulation of the EEP only holds when both matter systems and gravity are classical, and we do not know whether we should abandon or modify it when we consider quantum systems in a\,--\,possibly nonclassical\,--\,gravitational field. Here, we propose that the EEP is valid for a broader class of reference frames, namely Quantum Reference Frames (QRFs) associated to quantum systems. By imposing certain restrictions on the type of nonclassicality of the gravitational field, we develop a framework that enables us to formulate an extension of the EEP for such gravitational fields. This means that the EEP is valid in a much wider set of physical situations than what it is currently applied to, including those in which the gravitational field is in a quantum superposition state.
\end{abstract}
\maketitle

\section{Introduction}

In General Relativity, the metric acquires operational meaning via the geodesic motion of test particles, which act as probes and allow one to test the properties of the spacetime structure. The exact specification of geodesic motion, such as the fact that particles move along time-like or null geodesics, and that they can act as clocks ticking according to proper time, does not only rely on the field equations of the theory, but also needs a further independent condition. This condition consists in the requirement that the laws of physics reduce to those of special relativity in the vicinity of an arbitrary point in spacetime \cite{brown2016clarifying}, and is usually referred to as \emph{Einstein's Equivalence Principle} (EEP).

The EEP lies at the core of General Relativity (GR), and has deep implications both for the foundations of the theory and for its experimental verifications. On the conceptual side, it is possible to argue that confirmation of the EEP implies that gravity is a metric theory of spacetime~\cite{will_LR}. On the experimental side, the EEP constitutes a crucial tool to test possible modifications of GR~\cite{will_LR, tino2020precision}: an experimental violation would imply that a modified theory of gravity should be sought. 

All formulations of the EEP hold for purely classical matter and gravity. For situations in which quantum effects play a role at the interface between quantum theory and general relativity we do not know whether it is possible to retain the EEP, possibly in an extended form~\cite{aharonov1973quantum, lammerzahl1996equivalence, viola1997testing, rosi2017quantum, zych2018quantum, anastopoulos2018equivalence,seveso2017does, hardy2020implementation}, or we are forced to drastically modify our physical theories, as is the case for the principle of linear superposition in Penrose's state reduction~\cite{penrose1996gravity, penrose2014gravitization}.

There are two main ways in which quantum effects can modify the EEP. The first one, still in the regime of known physics, is the one of quantum systems in a classical gravitational field. In this case, there are conflicting opinions on whether the EEP can still be considered to hold, and of what its precise formulation would be~\cite{aharonov1973quantum, lammerzahl1996equivalence, viola1997testing, Davies_2004, rosi2017quantum, zych2018quantum, geiger2018proposal, anastopoulos2018equivalence,seveso2017does}. The second case, which is in the regime of unknown physics, is when spacetime is not classical. In this case, we do not know which physical laws to adopt, and which fundamental principles would lie at the basis of the theory.

We here take a first-principle approach to this question, and extend the EEP in a way that addresses both aspects of its generalisation to quantum theory. Specifically, both aspects follow from our proposal that the EEP is valid for a wider class of reference frames than those allowed in usual physical theories, i.e.\,reference frames associated with quantum systems. These correspond to Quantum Reference Frames (QRFs).

QRFs and quantum coordinate systems have been considered in quantum information, quantum foundations, and quantum gravity since 1967 \cite{aharonov1, aharonov2, aharonov3, dewitt1967quantum, brs_review, bartlett_communication, spekkens_resource, kitaev_superselection, palmer_changing, bartlett_degradation, smith_quantumrf, poulin_dynamics, skotiniotis_frameness, rovelli_quantum, poulin_reformulation, poulin_deformed, busch_relational_1, busch_relational_2, busch_relational_3, jacques, angelo_1, angelo_2, angelo_3, Giacomini:2017zju, perspective1, perspective2, Hardy:2018kbp, zych2018relativity,  hoehn2018switch, hohn2019switching, giacomini2019relativistic, hoehn2019trinity, castro2020quantum, hoehn2020equivalence, streiter2020relativistic, yang2020switching, de2020quantum, krumm2020}. In this work, we adopt the methods of the formalism introduced in Ref.~\cite{Giacomini:2017zju} and further explored in Refs.~\cite{perspective1, perspective2, giacomini2019relativistic, streiter2020relativistic, de2020quantum}. This approach gives a fully relational \cite{rovelli_relational} account of how physics is seen from the perspective of different QRFs. A conceptually similar approach has also been developed for switching between different quantum clocks \cite{hoehn2019trinity, hoehn2020equivalence}, also when they interact gravitationally \cite{castro2020quantum}. However, none of these works describe explicitly the gravitational degrees of freedom, and provide a fully covariant formulation of physics on curved spacetime.

We begin by considering a classical gravitational field, for which we identify the set of allowed transformations between two different QRFs as a set of ``quantum diffeormorphism transformations''. For all the cases that we consider, the structure of the allowed transformations corresponds to a quantum controlled-unitary operator, where the controlled operation is a classical diffeomorphism transformation and the QRF can be in an arbitrary quantum state. For instance, if the QRF is a quantum particle whose state is delocalised in a classical spacetime, the QRF transformation is a quantum superposition of classical diffeomorphism transformations, where the diffeomorphism transformation can be different for each position of the particle in spacetime.

Additionally, we allow for the possibility of having \emph{quantum superpositions of spacetimes}, which we define from first principles as a regime satisfying the following conditions:
\begin{itemize}[noitemsep,nolistsep]
	\item[\textbf{a.}] macroscopically distinguishable gravitational fields are assigned orthogonal quantum states. The gravitational fields are said to be distinguishable if they can be distinguished by measuring macroscopic observables;
	\item[\textbf{b.}] each well-defined gravitational field is described by general relativity;
	\item[\textbf{c.}] the quantum superposition principle holds for such gravitational fields.
\end{itemize}
In full generality, there is no accepted physical theory to describe this regime. However, there are physically concrete situations that satisfy the previous conditions, for example a massive body whose state is a quantum superposition of two semiclassical (i.e., coherent) states localised at two positions. An example of this is the gravitationally induced entanglement proposal~\cite{bose2017spin, marletto2017gravitationally}. Notice, however, that the superposed gravitational fields may but need not be diffeomorphically related to one another\footnote{Even if the two gravitational fields were diffeomorphically related, the two physical situations may not be equivalent because a of the presence of a test particle, which could become entangled with the gravitational field. The relevant physical information is in the coincidences in the two amplitudes between the particle and the field, and these are in general not diffeomorphically related.}. For each quantum state of gravity in the quantum superposition, corresponding to a classical spacetime, our formalism reduces to the description of the state of a quantum particle on curved spacetime. It follows from our assumptions \textbf{b.} and \textbf{c.} that all physical effects can be obtained by linearly superposing classical gravitational effects.

\begin{figure}\centering	 \large
\hfill 
\raisebox{-\height}{\textbf{a)}}
\raisebox{-\height}{\includegraphics[scale=0.5]{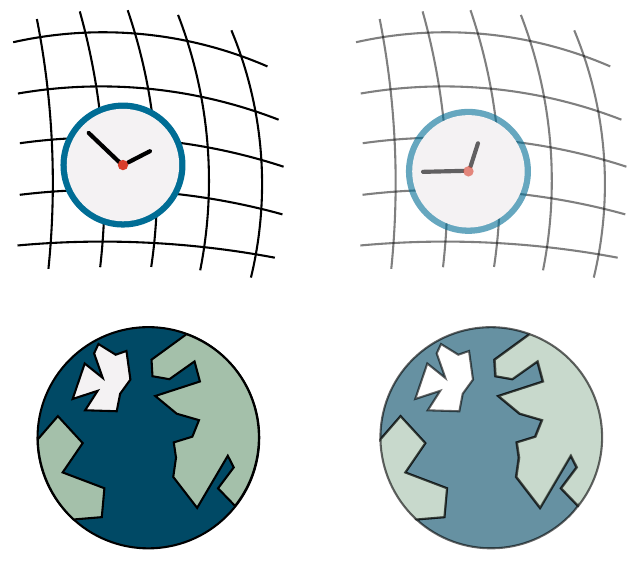}} \hfill
\raisebox{-\height}{\textbf{b)}}
\raisebox{-\height}{\includegraphics[scale=0.5]{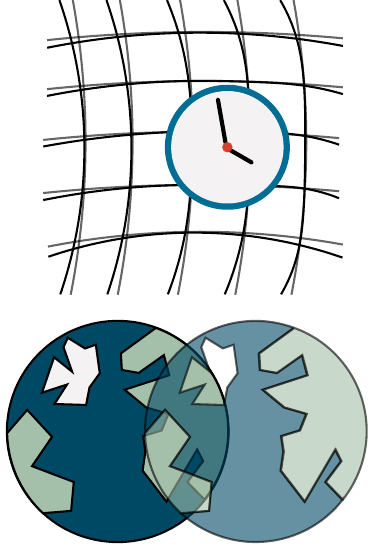}} \hfill\,\\
\caption{\label{fig:EEP clocks} Illustration of the generalisation of the Einstein Equivalence Principle (EEP) to quantum reference frames (QRFs). \textbf{a)} A mass in a spatial quantum superposition sources a superposition of classical gravitational fields. A clock in such superposition of gravitational fields, as seen from a general reference frame, is thus entangled with the gravitational field, and displays a different time according to the position of the mass. \textbf{b)} With a QRF transformation to the \emph{Quantum Locally Inertial Frame} (QLIF) of the clock, the metric which was originally in a superposition can be cast as a locally minkowskian metric at the location of the clock. The clock then ticks according to its proper time, which is well-defined, and decouples from the gravitational field, which can still be in a quantum superposition in the neighbourhood of the clock.}
\end{figure}

We consider general QRFs associated to quantum systems in classical spacetime and in a superposition thereof. With no ambition to construct a full theory of the superposition of gravitational fields, we build a minimal model that respects our conditions. In Sec.~\ref{sec:QLIFTransf}, we show that there always exists a unitary QRF transformation such that the gravitational field, and the superposition of gravitational fields, can be made locally minkowskian in the QRF associated to a quantum particle. This generalises the notion of Locally Inertial Frames (LIFs) to Quantum Locally Inertial Frames (QLIFs). We then formulate the following generalisation of the EEP by adapting the formulation of Ref.\,\cite{Misner:1974qy}
\begin{verse}
	\emph{In any and every \textbf{Quantum} Locally Inertial Frame (QLIF), anywhere and anytime in the universe, all the (nongravitational) laws of physics must take on their familiar non-relativistic form.} 
\end{verse}
A pictorial illustration of this principle, in a simplified setting, is provided in Fig.~\ref{fig:EEP clocks}.

In Sec.\,\ref{sec:FermiQRF} we extend the validity of our generalised EEP to the case of freely-falling frames in a superposition of spacetimes: we find a spacetime path-integral and fully covariant expression encoding the dynamics of a quantum particle in spacetime (and in a superposition thereof). As a consistency check, we recover the limit to a quantum particle in a Newtonian gravitational field in Sec.~\ref{sec:Newton}.

Notice that a Quantum Equivalence Principle has been proposed in Refs.~\cite{Hardy:2018kbp, hardy2020implementation}. There are conceptual similarities between this and our formulation of the EEP, but also important differences. The most relevant for this work is the invertibility of the transformations: while in our work the transformations, being unitary, are invertible, in Refs.~\cite{Hardy:2018kbp, hardy2020implementation} in general they are not. 

\section{Formalism}
\label{Sec:Formalism}

\begin{figure}
	\begin{center}
		\includegraphics[scale=0.7]{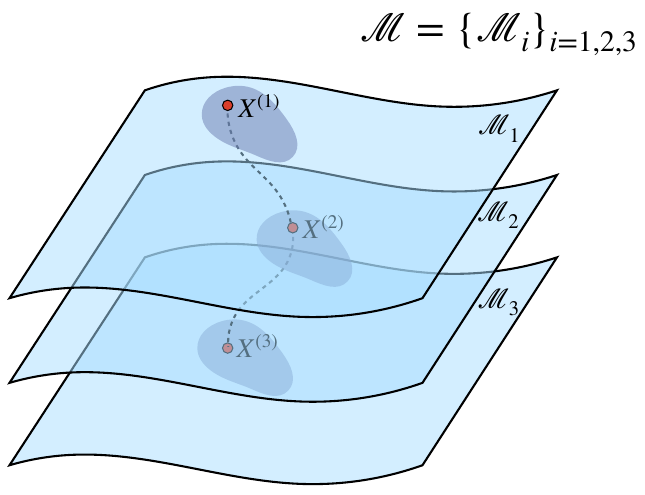}
		\caption{\label{fig:superpositionM} We formally describe the nonclassical structure of spacetime as a superposition of classical backgrounds enumerated as a set of manifolds $\mathcal{M} = \lbrace \mathcal{M}_i\rbrace_{i=1,\cdots, N}$ (with $N=3$ in the figure). On each classical manifold $\mathcal{M}_i$, we choose a coordinate system. Different points $X^{(i)}, \forall i=1, \cdots, N$ belonging to different manifolds are operationally identified with a ``physical point'' in the superposition of manifolds if a quantum system living in a superposition of those manifolds can be localised at those points $X^{(i)}$. In the figure, we represent with the shaded area the subregion of each classical manifold where the wavefunction of such quantum system has support. In the idealised case, the support of the quantum system reduces to a single point (red point in the figure) in every classical manifold. In the case considered, to each classical manifold $\mathcal{M}_i$ there corresponds a classical state of the gravitational field. In general, we do not know how to assign values to such gravitational field on a superposition of manifolds. We describe the gravitational field by correlating the spacetime point at which the field is defined with a quantum system, and we write this operation as $\sum_i \ket{g_i\triangleright \phi_i}$. This identification is natural, because tests of the gravitational field can be performed either directly, by measuring the gravitational radiation, or indirectly, as we choose here, via a probe particle. In the latter case, we can perform a measurement on the physical system to learn information about the gravitational field in the superposition of manifolds.}
	\end{center}
\end{figure}

In this section, we introduce a formalism to describe quantum systems living on a curved spacetime, and in a superposition of classical curved spacetimes. In all cases considered in this section, the quantum systems are considered as ``frozen'', i.e.\,not evolving dynamically. 

From now on, we consider two particles, $P$ and $M$, and the gravitational field, as described relative to a physical system $R$ (following the QRF formalism~\cite{Giacomini:2017zju}, $R$ is not explicitly described). The mathematical description of the two particles is the same, but we conceptually distinguish them: we assume that the particle $M$ is a probe particle (i.e., it does not backreact on the spacetime), but we do not make any assumption for particle $P$. The gravitational field corresponds to a superposition of classical (and macroscopically distinguishable) spacetimes. For simplicity, we take a finite set of dimension $N$, so that our set is described as $g = \lbrace g_i, \mathcal{M}_i \rbrace_{i=1, \cdots, N}$, where $g_i$ is a classical metric field defined on a differentiable manifold $\mathcal{M}_i$. On every manifold $i$, we assume the existence of a coordinate system, and we choose it to be centred in the position of $R$ (see Fig.~\ref{fig:superpositionM}). The origins of the coordinate systems in the different manifolds physically correspond to the position of $R$, and are hence identified. This specifies our choice of the initial \emph{quantum coordinate system} which we adopt. All our statements are not to be intended as valid globally, but only in a sufficiently small volume, in which the operators have support.

We now introduce an effective framework that satisfies our requirements \textbf{a.}-\textbf{c.}. Our assumptions \textbf{b.} and \textbf{c.} imply that there is a set of states $\ket{g^i}$, with $i=1, \cdots, N$, such that the gravitational field operator $\hat{g}_{\mu\nu}$ is diagonalised by such a set of states, and its eigenvalue corresponds to a classical configuration of the gravitational field. For instance, when the field operator is applied to a state $\ket{g^i}$ living on the manifold $\mathcal{M}_i$ labelled by $x$, we have $\hat{g}_{\mu\nu}(x)\ket{g^i} = g^i_{\mu\nu}(x)\ket{g^i}$. In addition, assumption \textbf{a.} implies that all states $\ket{g^i}$ are mutually orthogonal, i.e., $\braket{g^i|g^j}=\delta_{ij}$. These conditions together mean that the superposition of different manifolds is formally represented via a quantum superposition state of different gravitational fields, i.e.\,$\sum_{i=1}^N \alpha_i \ket{g^i}$, with $\alpha_i$ being complex coefficients.
Due to the orthogonality condition of the states of the gravitational fields, a unitary transformation on matter degrees of freedom acts as a quantum controlled unitary operation, namely it acts independently for each quantum state $\ket{g^i}$, and gravitational effects in each amplitude correspond to the usual general relativistic effects. The gravitational field operator, when evaluated on a state $\ket{g^i}$, transforms under a diffeomorphism transformation like in general relativity, with the quantum state being invariant under a diffeomorphism transformation.
 
To make sense of the identification of points across different manifolds, we use a localisation procedure via the coincidence between the gravitational field and a quantum system $M$. Specifically, we consider an \emph{operator-valued} gravitational field  $\hat{g}_{\mu\nu} (\hat{x}_M)$ (see, e.g., Refs.~\cite{stritzelberger2020coherent, Giacomini:2022hco} where the quantum field is evaluated at the position of the particle) such that
\begin{equation}
	\hat{g}_{\mu\nu} (\hat{x}_M) \ket{g^i}\ket{x^{(i)}}_M = g^i_{\mu\nu} (x^{(i)}) \ket{g^i}\ket{x^{(i)}}_M,
\end{equation}
where we have introduced a basis of the Hilbert space of $M$, $\ket{x^{(i)}}_M$. Here, $x^{(i)}$ labels the local coordinate system chosen on the manifold $\mathcal{M}_i$, and the index $i$ on the point reinforces that the manifold is only defined for the classical configuration of the gravitational field $g^i$. Notice that the eigenvalues of the position operators $\hat{x}_M^{(i)}$ have the same status as any abstract coordinate system in general relativity. The Hilbert space representation of the phase space operators $\hat{x}^{(i)}_M$ and $\hat{p}^{(i)}_M$, together with their generalisation to the superposition of manifolds, is described in App.~\ref{App:HilSpRep}. The eigenvalue $g^i_{\mu\nu} (x^{(i)})$ of the gravitational field operator is then naturally expressed in the local coordinate system $x^{(i)}$.

To formulate statements about the value that the components of the gravitational field take locally and identify points across different manifolds, we build a map which connects the points corresponding to the position of particle $M$ on different manifolds. This procedure assigns an operational meaning to the spacetime coordinates ---i.e., different spacetime coordinates of the particle in different manifolds all correspond to the same physical point as specified by the location of the particle, as illustrated in Fig.~\ref{fig:superpositionM}. This way of identifying points in the manifold is natural, because it corresponds to using a probe system as a measuring device for the gravitational field.

Concretely, this map assigns a value to the gravitational field not at an abstract spacetime point but at the location of a particle. In a quantum framework, since the particle can be delocalised over spacetime, this can be achieved by
evaluating the components of the gravitational field at every possible location of the probe particle $M$, i.e., at every point in the classical manifold $\mathcal{M}_i$ at which the quantum state of $M$ has support. This generalises the notion of coincidence in general relativity. Particle $M$ can be a general quantum system, to which we associate a state $\ket{\phi}_M \in \left\lbrace\mathcal{H}_i, \mathcal{M}_i, g_i \right\rbrace_{i=1, \cdots, N}$. Here, $\mathcal{H}_i$ is the Hilbert space corresponding to the classical manifold $\mathcal{M}_i$ with the classical metric field $g_i$. We assume that all $\mathcal{H}_i$, $i=1, \cdots N$ are isomorphic. The state $\ket{\phi}_M$ can be expressed in a coordinate representation only when we restrict it to a single manifold. We denote this operation as $\left. \ket{\phi}_M \right|_{\mathcal{M}_i} = \ket{\phi_i}_M$ and we write
\begin{equation}
	\ket{\phi_i}_M = \int d^4 x \sqrt{-g_i(x)} \phi_i(x) \ket{x^{(i)}}_M,
\end{equation}
where $d^4x \sqrt{-g_i(x)}$ is the covariant integration measure. On each classical manifold $\mathcal{M}_i$ we have the usual scalar product $\braket{x^{(i)}|x'^{(i)}}_M = \frac{\delta^{(4)}(x-x')}{\sqrt{-g_i(x)}}$. When we describe the state of both the gravitational field and the particle $M$ in a superposition of manifolds, the basis is given in terms of $\ket{x^{(i)}}_M\rightarrow \ket{g^i}\ket{x^{(i)}}_M$. In particular, the completeness and orthogonality condition is expressed as
\begin{equation}
	\frac{1}{4}\braket{g^i| g^j}\braket{x^{(i)}|x'^{(j)}}_M = \frac{\delta^{(4)}(x-x')}{\sqrt{-g_i(x)}}\delta_{ij},
\end{equation}
and the scalar product between basis elements defined on two different manifolds vanishes. The physical motivation for this requirement is that the gravitational fields defined on different manifolds are perfectly distinguishable, for instance because they correspond to different classical configurations of the mass sourcing the gravitational field.

With these definitions, we write a general state of the particle $M$ and of the gravitational field, restricted to the classical manifold $\mathcal{M}_i$, as
\begin{equation} \label{eq:gphi}
	\ket{g_i \triangleright \phi_i}=  \int d^4 x \sqrt{-g_i(x)} \phi_i(x) \ket{g^i } \ket{x^{(i)}}_M,
\end{equation}
where the symbol $\triangleright$ emphasises that the coordinate system chosen is valid within the restriction to the classical manifold $\mathcal{M}^i$ identified by the classical configuration $g^i$ of the gravitational field. The state defined in this way is normalised because we choose $\braket{\phi_i |\phi_i}=1$, and the scalar product between two different states is 
\begin{equation}
	\begin{split}
		\braket{g_i \triangleright \phi_i | g_i \triangleright \chi_i} &= \int d^4 x \sqrt{-g_i(x)}d^4 x' \sqrt{-g_i(x')}\phi^*_i(x') \chi_i(x)\braket{g^i | g^i}\braket{x'|x}_M=\\
		&= \int d^4 x \sqrt{-g_i(x)}\phi^*_i(x) \chi_i(x).
	\end{split}
\end{equation}

The state of Eq.~\eqref{eq:gphi} is formally similar to the \emph{spacetime smeared states} of Ref.~\cite{lake2019generalised}. However, while in  Ref.~\cite{lake2019generalised} these states represent superpositions of classical geometries, in our formalism the state of Eq.~\eqref{eq:gphi} represents a classical spacetime $g_i$ evaluated at the location of a quantum particle $M$, namely
\begin{equation}
	\hat{g}_{\mu\nu}(\hat{x}_M)\ket{g_i \triangleright \phi_i}=  \int d^4 x \sqrt{-g_i(x)} \phi_i(x)g^i_{\mu\nu}(x^{(i)})\ket{g^i } \ket{x^{(i)}}_M.
\end{equation}

We now extend the definition of the action of the gravitational field on a quantum state to the superposition of gravitational fields by taking the perspective of an observer who is not entangled with the gravitational field and the probe particle. For instance, the observer could be far-away from the region where the gravitational field and the particle are entangled\footnote{                   In the following, we show that it is always possible to choose such a disentangled observer even if the observer is located near the region where the gravitational field and the particle are entangled}. By linear superposition, we get
\begin{equation} \label{eq:ActionGravonM}
	\ket{g \triangleright \phi} = \sum_{i=1}^N c_i \ket{g_i \triangleright \phi_i} = \frac{1}{2}\sum_{i=1}^N c_i \int d^4 x \sqrt{-g_i(x)} \phi_i(x) \ket{g^i} \ket{x^{(i)}}_M,
\end{equation}
where the coefficients $c_i \in \mathbb{C}$ and the orthogonality condition for states living on different manifolds ensure that the full state is normalised. The square of the absolute value of the coefficients $c_i$, $|c_i|^2$, is the probability for the gravitational field to be found in the state $\ket{ g^i}$. The state of Eq.~\eqref{eq:ActionGravonM} should be distinguished from a mixture of classical spacetimes with probabilty $|c_i|^2$ for the gravitational field to be in the classical configuration $g_i$. The  relative phase between the $c_i$ coefficients is due to a quantum superposition of classical spacetimes and can be observed, for instance, in an interference experiment involving a massive body in a quantum superposition and a beam of massive particles scattered on it \cite{lindner2005testing}. Notice that an experiment of this sort, despite being currently out of reach with current technologies, is not irrealistic in the mid-term, as experiments can nowadays measure the gravitational field sourced by masses as light as $90 \,\text{mg}$ \cite{westphal2020measurement}.

\begin{figure}
	\begin{center}
		\includegraphics[scale=0.7]{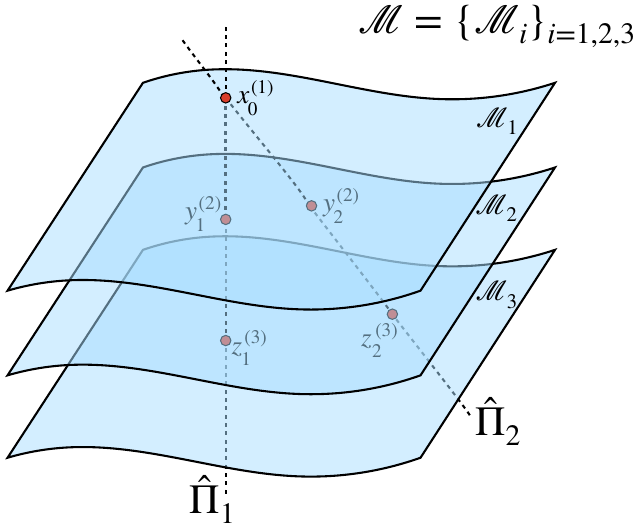}
		\caption{\label{fig:identification} Two different ways of identifying points across a superposition of classical manifolds. Using projector $\hat{\Pi}_1 = \ket{x_0^{(1)}}\bra{x_0^{(1)}} + \ket{y_1^{(2)}}\bra{y_1^{(2)}} + \ket{z_1^{(3)}}\bra{z_1^{(3)}}$, point $x_0$ in the first manifold $\mathcal{M}_1$ is identified with point $y_1$ in the manifold $\mathcal{M}_2$ and with point $z_1$ in the manifold $\mathcal{M}_3$. With the second projector $\hat{\Pi}_2 = \ket{x_0^{(1)}}\bra{x_0^{(1)}} + \ket{y_2^{(2)}}\bra{y_2^{(2)}} + \ket{z_2^{(3)}}\bra{z_2^{(3)}}$, point $x_0$ in the first manifold $\mathcal{M}_1$ is identified with point $y_2$ in the manifold $\mathcal{M}_2$ and with point $z_2$ in the manifold $\mathcal{M}_3$. The different physical identification of points corresponds to a different choice of measurement apparatus, where the projector is a mathematical abstraction representing the position of a measurement apparatus.}
	\end{center}
\end{figure}

The action of an arbitrary operator $\hat{O}$ on a state $\ket{g\triangleright \phi}$ is defined as a quantum controlled operator on each manifold $\mathcal{M}_i$, namely
\begin{equation}
	\hat{O} = \sum_{i=1}^N \hat{O}_i \ket{g^i}\bra{g^i}.
\end{equation} 
Hence, the operator $\hat{O}$ can be evaluated by linearly superposing the action of the restriction of the operator to each classical manifold $\left. \hat{O}\ket{\phi}_M \right|_{\mathcal{M}_i} = \hat{O}_i \ket{\phi_i}_M$, so that the full expression
\begin{equation} \label{eq:ControlOp}
	\hat{O}\ket{g \triangleright \phi}_M  = \sum_{i=1}^N c_i \hat{O}_i \ket{g_i \triangleright \phi_i}_M
\end{equation}
is equivalent to a controlled operator on the state of the gravitational field. Finally, in each classical manifold $\mathcal{M}_i$ we have
\begin{equation}
	\braket{x^{(i)}|p^{(i)}}_M = \frac{e^{\frac{i}{\hbar}p_\mu x^\mu}}{[-g_i(x)]^{1/4}},
\end{equation}
while the scalar product vanishes across different manifolds. In the previous equation, we have introduced the eingevector $\ket{p^{(i)}}_M$ of the momentum operators $\hat{p}_M^{(i)}$ that are canonically conjugated to the operators $\hat{x}_M^{(i)}$.

Our construction should be intended as valid locally. Specifically, in each classical spacetime labelled by the index $i$, we restrict to a region in which all the relevant operators have a converging Taylor series. In addition, we restrict the support of the wavefunctions to the region of convergence, and do not require the conditions on the unitarity and hermiticity of the operators to be valid with arbitrary precision, but only up to experimental resolution. Since all the operators are defined by their action on the superposition of spacetimes, these conditions are intended to be specified independently in each classical manifold.

We can now answer the question how the identification of points in different manifolds is performed by means of a projective measurement. This idea is represented in Fig.~\ref{fig:identification}, where two different choices of measurement apparatus are represented. In the first one, we identify point $x_0$ in the first manifold $\mathcal{M}_1$ with point $y_1 \in \mathcal{M}_2$ and with point $z_1 \in \mathcal{M}_3$ through the projector $\hat{\Pi}_1 = \ket{x_0^{(1)}}\bra{x_0^{(1)}} + \ket{y_1^{(2)}}\bra{y_1^{(2)}} + \ket{z_1^{(3)}}\bra{z_1^{(3)}}$. This situation physically corresponds to having a measurement apparatus which detects a quantum probe $M$ at its location in each classical manifold. Specifically, given a quantum state $\ket{g \triangleright \phi}_M = \sum_{i=1}^3 c_i \int d^4 x_i \sqrt{-g_i(x_i)} \phi_i(x_i) \ket{g^i} \ket{x_i^{(i)}}_M$, the application of the projector $\hat{\Pi}_1$ yields
\begin{equation}
		 \hat{\Pi}_1\ket{g \triangleright \phi}_M =  c_1 \phi_1(x_{0}) \ket{g^1}\ket{x_0^{(1)}}_M + c_2 \phi_2(y_{1}) \ket{g^2}\ket{y_1^{(2)}}_M + c_3 \phi_3(z_{1}) \ket{g^3}\ket{z_1^{(3)}}_M. 
\end{equation}
In particular, this means that the when the gravitational field is evaluated across the manifolds at the points which are identified by the projector, it does not need to take a single value, namely
\begin{equation}
	\begin{split}
		\hat{g}_{\mu\nu}(\hat{x}_M)\hat{\Pi}_1\ket{g \triangleright \phi}_M &=  c_1 g^1_{\mu\nu}(x_0) \phi_1(x_{0}) \ket{g^1}\ket{x_0^{(1)}}_M +\\
		&+ c_2 g^2_{\mu\nu}(y_1)\phi_2(y_{1}) \ket{g^2}\ket{y_1^{(2)}}_M + c_3 g^3_{\mu\nu}(z_1) \phi_3(z_{1}) \ket{g^3}\ket{z_1^{(3)}}_M.
	\end{split}
\end{equation}
In the next sections we make this statement precise via the definition of a QRF transformation in a superposition of spacetimes. If we choose a different record of the measurement apparatus, represented by a different projector $\hat{\Pi}_2$, i.e., $\hat{\Pi}_2 = \ket{x_0^{(1)}}\bra{x_0^{(1)}} + \ket{y_2^{(2)}}\bra{y_2^{(2)}} + \ket{z_2^{(3)}}\bra{z_2^{(3)}}$, we identify point $x_0 \in \mathcal{M}_1$ with point $y_2 \in \mathcal{M}_2$ and point $z_2 \in \mathcal{M}_3$. In this case, the projection operation yields a different identification, and we obtain
\begin{equation}
		\hat{\Pi}_2\ket{g \triangleright \phi}_M  =  c_1 \phi_1(x_{0}) \ket{g^1}\ket{x_0^{(1)}}_M + c_2 \phi_2(y_{2}) \ket{g^2}\ket{y_2^{(2)}}_M + c_3 \phi_3(z_{2}) \ket{g^3}\ket{z_2^{(3)}}_M. 
\end{equation}
Analogously to the previous case, we obtain
\begin{equation}
	\begin{split}
		\hat{g}_{\mu\nu}(\hat{x}_M)\hat{\Pi}_2\ket{g \triangleright \phi}_M &=  c_1 g^1_{\mu\nu}(x_0) \phi_1(x_{0}) \ket{g^1}\ket{x_0^{(1)}}_M +\\
		&+ c_2 g^2_{\mu\nu}(y_2)\phi_2(y_{2}) \ket{g^2}\ket{y_2^{(2)}}_M + c_3 g^3_{\mu\nu}(z_2) \phi_3(z_{2}) \ket{g^3}\ket{z_2^{(3)}}_M.
	\end{split}
\end{equation}
It is trivial that $\hat{g}_{\mu\nu}(\hat{x}_M)\hat{\Pi}_1\ket{g \triangleright \phi}_M  \neq \hat{g}_{\mu\nu}(\hat{x}_M)\hat{\Pi}_2\ket{g \triangleright \phi}_M  $. 

A different way of identifying points across different manifolds has been proposed in Ref.~\cite{Hardy:2018kbp} in the framework of quantum coordinate systems. There, the main difference to our approach is that the identification is not performed via a physical system. As a consequence, two points in two different manifolds which are identified in a specific quantum coordinate system are in general no longer identified after a quantum coordinate transformation.

\section{Transformation to a Quantum Locally Inertial Frame}
\label{sec:QLIFTransf}

In this section, we build the transformation to the Quantum Locally Inertial Frame (QLIF) centred in the position of a pointlike quantum particle having a state $\ket{\psi}_P$. The procedure we follow is a generalisation of the standard diffeomorphism transformation which defines a Locally Inertial Frame (LIF), which we review in App.\,\ref{App:LIFTrans}. Let us consider a transformation $x' \rightarrow \xi$, where the new coordinates are a function $\omega(x', x_P)$ of the old coordinates $x'$, with the change being centred about the point $x_P$, i.e.\,$\xi (x') = \omega (x', x_P)$. A classical gravitational field transforms as
\begin{equation}
	\tilde{g}_{\mu \nu} (\xi) = \Lambda^\alpha_\mu \Lambda^\beta_\nu g_{\alpha\beta} (x'(\xi)),
\end{equation}
where $\Lambda^\alpha_\mu = \frac{\partial x^{'\alpha}}{\partial \xi^\mu}$ and $x'(\xi)$ is the inverse coordinate transformation. The transformation can be Taylor-expanded about an arbitrary point $x$. It is sufficient to consider the second order expansion to identify a LIF. We define $b^\mu_\alpha =\frac{\partial \xi^\mu}{\partial x^{'\alpha}}\Big|_{x'=x}= b^\mu_\alpha (x)$ and $f_\mu^\alpha =\frac{\partial x^{'\alpha}}{\partial \xi^\mu}\Big|_{x'=x}= f_\mu^\alpha (x)$ the first derivative of the transformation and of its inverse evaluated at point $x$. To the leading order of approximation, $f_\mu^\alpha b^\mu_\beta = \delta^\alpha_\beta$ and $b^\mu_\alpha f_\nu^\alpha  = \delta^\mu_\nu$.

This diffeomorphism transformation can be realised on a classical spacetime and for a fixed $x$ with a unitary transformation $\hat{U}(\Lambda)\ket{x'}_M = \ket{\xi}_M$~\cite{DeWitt:1952js}. In a superposition of spacetimes indexed by the label $i$, we instead apply a different transformation $\Lambda^{(i)}(x)$ for each classical spacetime $g^i$ in the superposition, and for each position $x^{(i)}$ of the quantum system $P$ serving as the QRF. On a basis, the complete QRF transformation $\hat{S}$ corresponding to the change to a QLIF associated to the position of particle $P$ should act as
\begin{equation}
	\hat{S}\ket{x^{(i)}}_P \ket{g^i}\ket{{x'}^{(i)}}_M = \ket{-x^{(i)}}_R \ket{g^i}\ket{\xi^{(i)}(x)}_M,
\end{equation}
where the state of the gravitational field does not change because it is invariant under diffeomorphism transformations. This means that $\ket{g} = \ket{\tilde{g}}$ holds, with $\tilde{g}$ being the gravitational field after the transformation. The operator $\hat{S}$ preserves the scalar product between basis elements, thanks to the invertibility of a diffeomorphism transformation. It has the explicit form
\begin{equation} \label{eq:QRFGclass}
	\hat{S} = \mathcal{P}_{RP}\, \hat{U} \left( \Lambda^{(g)}(\hat{x}_P)\right),
\end{equation}
and it coincides with a standard transformation to a locally inertial frame when it is applied to a classical state of the gravitational field $\ket{g^i}$ and when particle $P$ is in a well-defined position in spacetime $x_P$. Here, $\mathcal{P}_{RP}$ is the so-called ``parity-swap'' operator acting as $\mathcal{P}_{RP}\ket{x}_P = \ket{-x}_R$, whose function is to map the relative position of $P$ to the origin of the initial reference frame $R$ to the relative position of $R$ to the origin of the final reference frame $P$ (see, e.g., Ref.~\cite{Giacomini:2017zju} for a detailed discussion of the role of this operator). The gravitational field operator, from our assumptions in Sec.~\ref{Sec:Formalism}, transforms as in classical GR when it is applied to the quantum state, namely
\begin{equation}
	\hat{S}\hat{g}_{\mu\nu}(\hat{x}_M) \ket{x}_P \ket{g^i}\ket{x'}_M = \Lambda^{(i)\alpha}_\mu (x) \Lambda^{(i)\beta}_\nu (x) \tilde{g}_{\alpha\beta}^i (\xi) \ket{-x}_R \ket{g^i}\ket{\xi}_M,
\end{equation}
where the inverse transformation has an analogous expression. In the most general case that we consider, our QRF transformation is then a \emph{controlled-unitary} transformation on the classical configuration of the gravitational field $\ket{g^i}$ and on the position of the QRF $\ket{x}_P$. We are then free to choose, for each $x$ and $g^i$, a different transformation which takes us to the LIF centred in $x$ on the spacetime $g^i$. Hence, by linearly extending the standard transformations to a LIF, we find that the QRF transformation $\hat{S}$ transforms to a set of coordinates which makes the metric minkowskian at the origin $x$ of the QRF, even when such an origin is in a quantum superposition of positions and in a superposition of gravitational fields from the initial perspective $R$. 

\subsection{Single classical spacetime (N=1)}
In the simplest case, we apply this transformation to the quantum state of the particle, the gravitational field, and the probe in the case where the spacetime is a classical background (and hence $N=1$). For later convenience, we explicitly include the state of the initial reference frame R, which is centred in the origin. The state in the QLIF of the particle $P$ is written as
\begin{equation}
	\begin{split}
			&\hat{S}\ket{\psi ; g\triangleright\phi} \ket{0}_R = \int d^4 x \sqrt{-g(x)} \psi(x) \int d^4 x' \sqrt{-g(x')} \phi (x') \ket{g} \ket{\xi(x')}_M \ket{-x}_R \ket{0}_P=\\
		=& \int d^4 x \sqrt{-g(x)} \psi(x) \int d^4 \xi \sqrt{-\tilde{g}(\xi)} \phi (x'(\xi))\ket{\tilde{g}} \ket{\xi}_M \ket{-x}_R \ket{0}_P,
	\end{split}
\end{equation}
where we have used that $|det(\Lambda)| \sqrt{-g(x'(\xi))} = \sqrt{-\tilde{g}(\xi)}$.

We can now evaluate the gravitational field $\hat{g}_{\mu\nu}(\hat{x}_P)$ on the quantum state as
\begin{equation}
	\hat{g}_{\mu\nu}(\hat{x}_P) \hat{S} \ket{\psi}_P \ket{g\triangleright\phi} \ket{0}_R = \tilde{g}_{\mu\nu}(0)\hat{S} \ket{\psi}_P \ket{g\triangleright\phi} \ket{0}_R,
\end{equation}
where $\tilde{g}_{\mu\nu}(0) = f^\alpha_\mu(x) f^\beta_\nu(x) g_{\alpha\beta}(x) = \eta_{\mu\nu}$ thanks to our definition of the QRF transformation. With a similar procedure we could also verify that the first derivative of the gravitational field vanishes locally, at the origin of the QRF, as required by the EEP.

\subsection{Superposition of classical spacetimes ($N > 1$)}
\label{sec:QLIFinSupSpacetime}

When we consider a quantum superposition of spacetimes, the transformation to the QLIF works similarly to the case $N=1$ but, in addition, it depends on the metric field $g_i$ for different values of $i$. By linear superposition, and repeating similar steps to the previous case, we find 
\begin{equation}
	\begin{split}
		&\hat{S}\ket{\Psi}\ket{0}_R =\frac{1}{\sqrt{N}} \sum_{i=1}^N \hat{S}_i\ket{\psi_i}\ket{g_i \triangleright \phi_i}\ket{0}_R=\\
		&=\frac{1}{\sqrt{N}}\sum_{i=1}^N \int d^4 x \sqrt{-g_i(x)} \psi_i(x)\int d^4 \xi \sqrt{-\tilde{g}_i(\xi)} \phi_i(x'(\xi^{(i)}))\ket{\tilde{g}^i}\ket{\xi^{(i)}}_M \ket{-x^{(i)}}_R \ket{0^{(i)}}_P.
	\end{split}
\end{equation}
Since $\hat{S}$ acts as a controlled unitary transformation $\hat{S}_i$ on a set of orthogonal states, the transformation $\hat{S}$ is unitary. Evaluating the gravitational field at the origin of the QRF, we find
\begin{equation}
	\hat{g}_{\mu\nu}(\hat{x}_P) \hat{S} \ket{\Psi} \ket{0}_R = \tilde{g}^i_{\mu\nu}(0)\hat{S} \ket{\Psi} \ket{0}_R,
\end{equation}
with $\tilde{g}^i_{\mu\nu}(0)= (f^i)^\alpha_\mu(x) (f^i)^\beta_\nu(x) g_{\alpha\beta}(x) = \eta_{\mu\nu}$ for all values of $i$. Analogously to the case $N=1$, also for $N>1$ the QRF transformation can be chosen so that the first derivative of the gravitational field vanishes locally, at the origin of the QRF. This means that, in a quantum superposition of spacetimes it is also possible to find a QLIF which is locally minkowskian. 

We have thus shown that it is possible to build a QRF transformation to a QLIF, in which the metric looks locally minkowskian even when, in the initial frame, the gravitational field was in a superposition of classical spacetimes. Importantly, our result shows that the metric is locally minkowskian in the same sense as in classical GR: we have applied a classical diffeomorphism on a metric field $g^i$, but in a quantum superposition controlled on the point $x^{(i)}$ corresponding to the centre of the QLIF and by the manifold $i$, on each of which a different gravitational field is defined. Hence, this result is a generalisation of the EEP in the presence of a superposition of geometries.

\section{Quantum free-falling particle in a classical gravitational field}
\label{sec:FermiQRF}

Let us consider a classical point particle on a curved classical background, which is in free fall in the gravitational field. Technically, this means that the particle obeys the geodesic equation
\begin{equation}
	\frac{d^2 x^\mu}{d\tau^2} + \Gamma^\mu_{\alpha\beta} \frac{d x^\alpha}{d\tau}\frac{d x^\beta}{d\tau}=0.
\end{equation}
In classical general relativity, a particle in free fall can be described with an action principle, where the action is $S= \int d\tau \sqrt{g_{\mu\nu} \dot{x}^\mu \dot{x}^\nu}$. The corresponding dispersion relation in momentum space (i.e., the Klein-Gordon equation on curved background) is $C = g_{\mu\nu} p^\mu p^\nu - m^2 c^2$. Let us define $\hat{C} = \mathcal{W}(g_{\mu\nu} p^\mu p^\nu - m^2 c^2)$ as the Weyl-ordered operator corresponding to the classical dispersion relation, where $\mathcal{W}$ stands for the choice of the ordering of the operators. Then, we write the physical state (i.e., the state satisfying the physical constraint) as
\begin{equation}
	\ket{\psi_p} = \frac{1}{2\pi \hbar} \int d\tau e^{-\frac{i}{\hbar} \hat{C}\tau} \int d^4 x_0 \sqrt{-g(x_0)} \psi_k(x_0) \ket{x_0}_P,
\end{equation}
where $\ket{\psi_p}$ is the state on the Physical Hilbert space (i.e., the state that satisfies the constraint) and $\ket{\psi_k} = \int d^4 x_0 \sqrt{-g(x_0)} \psi_k(x_0) \ket{x_0}_P$ is the state on the Kinematical Hilbert space (i.e., it does not need to satisfy the constraint). The state $\ket{\psi_p}$ can be recast as a path-integral (see App. \ref{App:PathIntegral} for details and Refs.~\cite{dewitt1957dynamical, fradkin1991path, padmanabhan1994path, poisson2011motion, alsing2001phase} for alternative approaches to the quantisation of the path integral of a relativistic particle)
\begin{equation} \label{eq:actionPhysCurved}
		\ket{\psi_p} = \frac{1}{2\pi \hbar} \int d\tau \int d\mu[x_0] d\mu[x(\tau)]  K(x_0, x(\tau), 0, \tau)  \psi_k(x_0) \ket{x(\tau)},
\end{equation}
where $d\mu(x) = \sqrt{-g(x)}d^4x$ and
\begin{equation}
	\begin{split}
		 & K(x_0, x(\tau), 0, \tau) =  [-g(x_0)]^{-1/4}[-g(x(\tau))]^{-1/4}\int \mathcal{D}[x(\tau)]  e^{\frac{i}{\hbar} \int_0^\tau ds \mathcal{L}(x(s), \dot{x}(s), s)},
	\end{split}
\end{equation}
where $\mathcal{L}(x(s), \dot{x}(s), s) = \sqrt{g_{\mu\nu} (x(s)) \dot{x}^\mu \dot{x}^\nu} $. This means that we can write the physical state of a quantum particle in a classical gravitational field as the (infinite) sum of all possible \emph{classical} trajectories with a relative phase corresponding to the classical action. For compactness of notation, we write the state that we obtained as
\begin{equation}
	\ket{\psi_p} = \int d\tau \omega(x (\tau)) \ket{x(\tau)}
\end{equation}
where $\omega(x (\tau))= \frac{1}{2\pi \hbar} \int d\mu[x_0] d\mu[x(\tau)] K(x_0, x(\tau), 0, \tau)  \psi_k(x_0) $. This can be seen as a ``history state''\footnote{Notice that this is a different notion to other ``history states'' in the literature, e.g. those considered in Refs.~\cite{griffiths1993consistent, halliwell1994review, cotler2016entangled}.} for a quantum particle in a gravitational field, i.e., the state $\ket{x(\tau)}$ evolves in $\tau$ according to the classical geodesic equation (see App.~\ref{App:HistoryState}).

\subsection{Transformation to the QRF of a free-falling particle in a classical gravitational field}

Let us start by considering the total state of a particle, which does not backreact on the gravitational field, and of the gravitational field 
\begin{equation}
		\ket{\psi_p; g\triangleright \phi} \propto \int d\tau \omega(x (\tau)) \ket{x(\tau)}_P\int d^4x' \sqrt{-g(x')}\phi(x')\ket{g}\ket{x'}_M \ket{0}_R.
\end{equation}
In the initial frame $R$, when we evaluate the state of the gravitational field at the location of the particle, we find 
\begin{equation}
	\hat{g}_{\mu\nu}(\hat{x}_P)\ket{\psi_p; g\triangleright \phi} \propto \int d\tau g_{\mu\nu}(x(\tau)) \omega(x (\tau)) \ket{x(\tau)}_P\int d^4x' \sqrt{-g(x')}\phi(x')\ket{g}\ket{x'}_M \ket{0}_R.
\end{equation}
which means that the value of the gravitational field is correlated with the position of the freely falling particle along the entire worldline.

If we consider the QRF transformation from the previous section, and performing analogous calculations, the transformed state becomes
\begin{equation}
	\begin{split}
		\hat{S}\ket{\psi_p; g\triangleright \phi}  &\propto \int d\tau \omega(x (\tau)) \ket{-x(\tau)}_R \int d^4\xi \sqrt{-\tilde{g}(\xi)}\phi(x'(\xi))\ket{\tilde{g}}\ket{\xi}_M \ket{0}_P,
		\end{split}
\end{equation}
where $\tilde{g}_{\mu\nu}(\xi) = \Lambda^\alpha_\mu(x(\tau)) \Lambda^\beta_\nu(x(\tau)) g_{\alpha\beta}(x'(\xi))$ and the transformation $\xi (x') = \omega(x', x(\tau))$ depends on the dynamical position $x(\tau)$ of particle $P$. In the final QRF, the evaluation of the gravitational field at an arbitrary point along the geodesic yields
\begin{equation}
	\hat{g}_{\mu\nu}(\hat{x}_P)\ket{\tilde{\Psi}}^{(P)}= g_{\mu\nu}(0)\ket{\tilde{\Psi}}^{(P)},
\end{equation}
where $g_{\mu\nu}(0)= \eta_{\mu\nu}$ with a similar reasoning to the non dynamical case. This result implies that the field is locally minkowskian in the QRF of a quantum particle in free fall in a classical gravitational field.

\subsection{Transformation to the QRF of a free-falling particle in a superposition of gravitational fields}

We now generalise the previous result to a superposition of gravitational fields. The method is completely analogous to the one we have used in Sec.~\ref{sec:QLIFinSupSpacetime} to define the QLIF of a static particle in a superposition of spacetimes. In the present case, the full state is
\begin{equation}
	\ket{\Psi} = \sum_{i=1}^N c_i \ket{\psi_{p,i}}\ket{ g_i \triangleright \phi_i}\ket{0}_R,
\end{equation}
where $c_i$ are arbitrary coefficients such that $|c_i|^2$ is the probability of the gravitational field to be in the classical configuration $g^i$, and each $\ket{\psi_{p,i}}$ is the physical state solving the constraint
\begin{equation}
	\ket{\psi_{p,i}} = \frac{1}{2\pi \hbar} \int d\tau e^{-\frac{i}{\hbar} \hat{C}^i\tau} \int d^4 x_0 \sqrt{-g_i(x_0)} \psi_{k,i}(x^{(i)}_0) \ket{x^{(i)}_0}_P,
\end{equation}
with $\psi_{k,i}(x^{(i)}_0)$ being the kinematical state and $\hat{C}^i = \mathcal{W}(g^i_{\mu\nu} p^\mu p^\nu - m^2 c^4)$ the Weyl-ordered operator with metric field $g^i$. We can thus write
\begin{equation}
		\ket{\Psi} =\sum_{i=1}^N  c_i\int d\tau \omega_i(x (\tau)) \ket{x(\tau)^{(i)}}_P \int d^4x \sqrt{-g_i(x)}\phi_i(x)\ket{g^i}\ket{x^{(i)}}_M \ket{0}_R,
\end{equation}
where $\omega_i(x(\tau)) = \frac{1}{2\pi\hbar}\int d^4 x_0 d^4 x(\tau) [-g_i (x_0)]^{1/4}[-g_i (x(\tau))]^{1/4} \int \mathcal{D}_i [x(\tau)] e^{\frac{i}{\hbar}\int_0^\tau ds \mathcal{L}_i (x(s), \dot{x}(s), s)}\psi_{k,i}(x_0)$ and the Lagrangian $\mathcal{L}_i (x(s), \dot{x}(s), s) = \sqrt{g^i_{\mu\nu}(x(s))\dot{x}^\mu \dot{x}^\nu}$ (we omit here the $i$ index on the coordinates to avoid overloading the notation).

Thanks to the linearity of the QRF transformation $\hat{S}$ and to the mutual orthogonality of the states $\ket{g^i}$ for different values of $i$, we can linearly superpose the results of the previous section, by letting the transformation $\hat{U}_R$ from Eq.~\eqref{eq:QRFGclass} be controlled by the configuration $i$ of the gravitational field $g_i$. Hence, we obtain $\ket{\tilde{\Psi}}^{(P)} = \hat{S}\ket{\Psi}$, where
\begin{equation}
		\ket{\tilde{\Psi}}^{(P)} = \sum_{i=1}^N  c_i\int d\tau \omega_i(x (\tau)) \ket{-x(\tau)^{(i)}}_R \int d^4\xi \sqrt{-\tilde{g}_i(\xi)}\phi_i(x'(\xi))\ket{\tilde{g}^i}\ket{\xi^{(i)}}_M \ket{0}_P,
\end{equation}
where $\tilde{g}^i_{\mu\nu} (\xi^{(i)}) = \Lambda^{(i)\alpha}_\mu (x^{(i)}(\tau)) \Lambda^{(i)\beta}_\nu (x^{(i)}(\tau)) g^i_{\alpha\beta}(x'(\xi))$. We can then choose the coefficients $ f^i (x(\tau))$, for each $i=1, \cdots, N$ and for each value of the parameter $\tau$, such that $\tilde{g}^i(0) = f^i (x^{(i)}(\tau)) f^i (x^{(i)}(\tau)) g^i[x^{(i)}(\tau)] = \eta$. Evaluating the gravitational field at the origin of the final QRF $P$, we obtain
\begin{equation}
	\hat{g}_{\mu\nu}(\hat{x}_P)\ket{\tilde{\Psi}}^{(P)}= g^i_{\mu\nu}(0) \ket{\tilde{\Psi}}^{(P)},
\end{equation}
where $g^i_{\mu\nu}(0)= \eta_{\mu\nu}$ for all values of $i$. This result shows that the field is locally minkowskian in the QRF of a quantum particle in free fall in a superposition of gravitational fields along the entire trajectories.

\section{The Newtonian limit: recovering a background time}
\label{sec:Newton}

As a consistency check of the validity of our description, we show how to recover the Hamiltonian picture for a quantum particle in a Newtonian gravitational field. The Newtonian limit of the quantum particle in free falling in a gravitational background is obtained when the metric is taken to be, apart for higher-order terms in the magnitude of the field
\begin{equation} \label{eq:Newtong}
	\begin{split}
		& g_{00} = 1 + \frac{2\Phi(\mathbf{x})}{c^2};\\
		& g_{0i} = 0;\\
		& g_{ij} = -\delta_{ij},
	\end{split}
\end{equation}
where $\Phi(\mathbf{x}) = - \frac{G M_{\odot}}{|\mathbf{x}-R_{\odot}|}$ is the Newtonian potential due to a system with mass $M_{\odot}$ at position $R_{\odot}$ from the point of view of the initial reference frame. Notice that the potential only depends on the spatial coordinates.

By requiring the positive energy condition in the Newtonian limit, i.e. that the constraint be modified to 
\begin{equation}
	\hat{C}\theta(\hat{p}_0) = \left[\sqrt{g^{00}}\hat{p}_0 - \sqrt{|\hat{\mathbf{p}}|^2 + m^2 c^2}\right]\left[\sqrt{g^{00}}\hat{p}_0 + \sqrt{|\hat{\mathbf{p}}|^2 + m^2 c^2}\right]\theta(\hat{p}_0),
\end{equation}
where $\theta(X)$ is the step function defined as $\theta(X)= 1$ for $X\geq 0$, $\theta(X)= 0$ otherwise, we find that the only part of the constraint which can be annihilated by the state of the particle $P$ is
\begin{equation}
	\hat{C}' = \sqrt{g^{00}}\hat{p}_0 - \sqrt{|\hat{\mathbf{p}}|^2 + m^2 c^2}.
\end{equation}
For later convenience, we also introduce the internal degrees of freedom of the particle, which we will use as an internal clock to test the Equivalence Principle in a concrete setup. Knowing that, due to the mass-energy equivalence, the internal energy contributes to the total mass, we can replace $m$ in the previous equation with $m \rightarrow m + \frac{\hat{H}_I}{c^2}$. If we expand $\hat{C}'$ in a Taylor series, we obtain 
\begin{equation}
	\hat{C}' = \sqrt{g^{00}}\hat{p}_0 - mc - \frac{|\hat{\mathbf{p}}|^2}{2m c} + \frac{|\hat{\mathbf{p}}|^4}{8 m^3 c^3} - \frac{\hat{H}_I}{c}\left( 1 - \frac{|\hat{\mathbf{p}}|^2}{2m^2 c^2} \right).
\end{equation}
We now condition on the time coordinate $t$ as measured in the initial reference frame (corresponding to the laboratory reference frame inertial with the Earth). Knowing that $\bra{x^0 = ct} \hat{C}' \ket{\psi_p}=0$, knowing that $\hat{p}_0 \ket{x} = \frac{i \hbar}{c} \frac{d}{ dt}\ket{x}$, and defining $\ket{\psi_t}= \braket{ct | \psi_p}$, we obtain
\begin{equation}
	\sqrt{g^{00}} i\hbar \frac{d \ket{\psi_t}}{dt} = \left[ mc^2 + \frac{|\hat{\mathbf{p}}|^2}{2m} - \frac{|\hat{\mathbf{p}}|^4}{8 m^3 c^2} + \hat{H}_I\left( 1 - \frac{|\hat{\mathbf{p}}|^2}{2m^2 c^2} \right)\right] \ket{\psi_t}.
\end{equation}
In order to find the usual Schr{\"o}dinger equation of a quantum particle in a Newtonian field, we need to invert the operator $\sqrt{g^{00}(\hat{\mathbf{x}})} = \sqrt{1 - \frac{\Phi(\hat{\mathbf{x}})}{c^2}}$. In the weak field limit the square root is always positive when applied to any quantum state, hence we find 
\begin{equation}
	i\hbar \frac{d \ket{\psi_t}}{dt} = \left[ mc^2 + \frac{|\hat{\mathbf{p}}|^2}{2m} + m \Phi(\hat{\mathbf{x}}) - \frac{|\hat{\mathbf{p}}|^4}{8 m^3 c^2} +\Phi(\hat{\mathbf{x}}) \frac{|\hat{\mathbf{p}}|^2}{2m c^2}  + \hat{H}_I\left( 1 + \frac{\Phi(\hat{\mathbf{x}})}{c^2}- \frac{|\hat{\mathbf{p}}|^2}{2m^2 c^2} \right)\right] \ket{\psi_t}.
\end{equation}
If we take the particle to move with velocities which are slow compared to the speed of light, we can neglect the relativistic terms and only keep the gravitational ones. We thus find
\begin{equation}
	i\hbar \frac{d \ket{\psi_t}}{dt} = \left[ mc^2 + \frac{|\hat{\mathbf{p}}|^2}{2m} + m \Phi(\hat{\mathbf{x}}) + \hat{H}_I\left( 1 + \frac{\Phi(\hat{\mathbf{x}})}{c^2} \right)\right] \ket{\psi_t} = \hat{H}_{PI}\ket{\psi_t}.
\end{equation}
From the last equation, we see that the particle can be described in the background time of the reference frame as if it was in an external potential, and with the internal clock of the particle (i.e., the internal degrees of freedom) ticking according to the proper time. Proper time is related to time $t$ by a factor $\left( 1 + \frac{\Phi(\hat{\mathbf{x}})}{c^2} \right)$, corresponding to gravitational time dilation.

We now use this result to write the full state of the particle by constructing a ``history state'' with respect to a clock in the laboratory frame
\begin{equation} \label{eq:Newton1}
	\ket{\Psi} = \int dt e^{- \frac{i}{\hbar}\hat{H}_{PI} t} \ket{\psi_0}^{(L)} \ket{t}_L \ket{\phi \triangleright g},
\end{equation}
where $\ket{\psi_0}^{(L)} = \int dE_I d^3 x \psi_0 (\mathbf{x}, E_I) \ket{\mathbf{x}, E_I}$ is formally analogous to the initial state of the particle\footnote{Notice that $\ket{\psi_0}^{(L)}$ does not have the operational meaning of an initial state, since the expression is integrated over all possible times of the laboratory clock.}, $\ket{E_I}$ is the eigenstate of the Hamiltonian $\hat{H}_I$, and $\ket{\phi \triangleright g}$ is the state of the probe and of the Newtonian gravitational field in Eq.~\eqref{eq:Newtong}. We now define the Hamiltonian of the laboratory clock to be $\hat{H}_L = \int dE_L E_L \ket{E_L}_L \bra{E_L}$, with $\braket{t|E_L} = \frac{1}{\sqrt{2\pi\hbar}} e^{\frac{i}{\hbar}E_L t}$.  We first notice that the dynamical part of the quantum state in Eq.~\eqref{eq:Newton1} is equivalent to
\begin{equation} \label{eq:Newton2}
		\int dt e^{- \frac{i}{\hbar}\hat{H}_{PI} t} \ket{\psi_0}^{(L)} \ket{t}_L =  \int d\tau  e^{- \frac{i}{\hbar}\hat{H}'_{PL} \tau}\ket{\psi_0}^{(I)}\ket{\tau}_I,
\end{equation}
where we have used that $\left[\frac{\Phi(\hat{\mathbf{x}})}{c^2}, \frac{|\hat{\mathbf{p}}|^2}{2m}\right]$ is higher-order in our approximation and we have defined the new variable $\tau = \left( 1 + \frac{\Phi(\mathbf{x})}{c^2} \right)t$ as the proper time in the particle's rest frame P (see App.~\ref{App:NewtonstateEquiv} for the proof of the equivalence in Eq.~\eqref{eq:Newton2}). We also have 
\begin{align}
	&\hat{H}'_{PL} = \left[ mc^2 + \frac{|\hat{\mathbf{p}}|^2}{2m} +\left( 1 - \frac{\Phi(\hat{\mathbf{x}})}{c^2} \right) \hat{H}_L\right],\\
	&\ket{\psi_0}^{(I)}= \frac{1}{2\pi\hbar} \int d\tau_0 d^3 x dE_I dE_L \left( 1 - \frac{\Phi(\mathbf{x})}{c^2} \right)\psi_0 (\mathbf{x}, E_I) e^{\frac{i}{\hbar}(\hat{H}'_{PL}+ E_I )\tau_0} \ket{\mathbf{x}}_P \ket{E_L}_L.
\end{align} 
We can now apply the transformation to the QLIF of particle P from Eq.~\eqref{eq:QRFGclass} and evaluate the gravitational field at the origin of the QRF. Notice that, since we are explicitly describing the initial and final reference frames of the two clocks, in this case the parity-swap operator should not be included in the transformation. We find
\begin{equation} \label{eq:Newton3}
	\hat{g}_{\mu\nu}(\hat{x}_P)\hat{S} \ket{\Psi}= \eta_{\mu\nu}\hat{S} \ket{\Psi}.
\end{equation}
We thus find that the full state can either be described from the point of view of the laboratory as a particle falling in the gravitational field and carrying a gravitationally shifted clock by the factor $\left( 1 + \frac{\Phi(\hat{\mathbf{x}})}{c^2}\right)$, as in Eq.~\eqref{eq:Newton1}, or as a free-particle in the rest frame of the particle clock,  where the clock of the laboratory appears gravitationally shifted with the inverse factor $\left( 1 - \frac{\Phi(\hat{\mathbf{x}})}{c^2}\right)$ and the metric is locally minkowskian, as described in Eq.~\eqref{eq:Newton3}. 

It is straightforward to extend these considerations to superpositions of Newtonian gravitational fields. These two descriptions are governed by the two Hamiltonian operators $\hat{H}_{PI}$ and $\hat{H}'_{PL}$ respectively. Knowing that the Einstein Equivalence Principle, according to Schiff conjecture, can be experimentally verified by testing the Weak Equivalence Principle and the Universality of the Gravitational Redshift, an interferometric experiment of the type proposed in Ref.~\cite{zych_interferometric}, where the clock moves in a superposition of trajectories in a classical gravitational field, would be sufficient to test the EEP in QRFs. If, in addition, the gravitational field itself were in a superposition, this would allow for a test of the EEP for such superpositions.

\section{Conclusions}

In this work we proposed that the Einstein Equivalence Principle holds for a more general class of reference frames, namely quantum reference frames. This generalisation allows us to extend at once the validity of the Equivalence Principle to both quantum systems that are delocalised in a classical spacetime, and to a quantum superposition of spacetimes, a regime for which there is currently no established physical theory. Key to the generalisation is the definition of a transformation to the quantum reference frame associated to the position of a particle, which can be entangled with the different configurations of the gravitational field. This defines a \textbf{Quantum Locally Inertial Frame}.

In addition, we have found a spacetime path-integral description encoding the dynamics of a quantum particle, and shown that the state of a freely-falling quantum particle can be expressed as an infinite sum over all possible classical trajectories.

We have further shown that, even in the case when the quantum particle is evolving dynamically, we can find a quantum reference frame transformation to the freely-falling frame of the particle, such that the metric looks locally minkowskian, both in a classical curved spacetime and in a superposition thereof along the entire trajectory of the particle.

In our work, we have considered a superposition of classical manifolds, and we have shown that, in a Quantum Locally Inertial Frame, different manifolds can be suitably identified, locally, to a single Minkowski metric structure. Our approach has the advantage that it enables us to take the ``internal'' perspective of a quantum particle. In future work, this feature will be used to describe the gravitational field in a region, and not only locally at the origin of the Quantum Locally Inertial Frame. This result suggests that the partition into manifolds is not a perspective-independent feature, but it depends on the quantum reference frame. A consequence of this fact is that it is not necessary to consider a far-away observer to have a description that is not influenced, locally, by the quantum superposition of spacetime structures: it is always possible to define a local quantum reference frame in which the metric decouples, locally, from all other physical systems. In this sense, any Quantum Locally Inertial Frame associated to a physical system satisfies this requirement. A similar result was derived in Ref.~\cite{castro2020quantum} for clocks which get entangled via the gravitational interaction. In that case, however, the full metric structure of the theory was not considered.

These results extend the Einstein Equivalence Principle to quantum reference frames in a superposition of gravitational fields, and hence suggest that this principle could be a guide towards a formulation of a theory at the interface between quantum theory and gravity in a more general setting than what is usually understood. Historically, the Einstein Equivalence Principle played a pivotal role in the development of the theory of General Relativity, and is one of the preferential tools to test the theoretical underpinnings of the theory, as it arguably encodes the metric structure of the theory. One may suspect that its generalisations to quantum reference frames and nonclassical spacetimes might play an equally important role in guiding the research to developing a unifying framework of quantum theory and gravity.

\acknowledgments{\v{C}. B. would like to thank Esteban Castro Ruiz, Anne-Catherine de la Hamette, and Viktoria Kabel for helpful discussions. F.G. acknowledges support from Perimeter Institute for Theoretical Physics.
Research at Perimeter Institute is supported in part by the Government of Canada through the
Department of Innovation, Science and Industry Canada and by the Province of Ontario through
the Ministry of Colleges and Universities. {\v C}. B. acknowlegdes support from the research platform TURIS, from the European Commission via Testing the Large-Scale Limit of Quantum Mechanics (TEQ)(No.   766900) project, and  from  the  Austrian-Serbian bilateral scientific cooperation  no. 451-03-02141/2017-09/02, and by the Austrian Science Fund (FWF) through the  SFB  project  BeyondC and a grant from the Foundational Questions  Institute (FQXi) Fund.   This  publication  was  made  possible through the  support  of  the ID 61466 grant from the John Templeton Foundation, as part of The Quantum Information Structure of Spacetime (QISS) Project (qiss.fr). The opinions expressed in this publication are those of the author(s) and do not necessarily reflect the views of the John Templeton Foundation.}

\appendix

\section{Hilbert space representation of phase space operators in a superposition of manifolds}
\label{App:HilSpRep}

We introduce the phase-space operators $\hat{x}^\mu$, $\hat{p}_\nu$ (for simplicity of notation, we here omit the subscript $X$ referring to the quantum system) living in the superposition of manifolds, together with their restriction to the manifold $\mathcal{M}_i$, $\left. \hat{x}^\mu \right|_{\mathcal{M}_i} = \hat{x}^\mu_{(i)}$ and $\left. \hat{p}_\nu \right|_{\mathcal{M}_i} = \hat{p}_\nu^{(i)}$. In every classical manifold $\mathcal{M}_i$, the operators have the following Hilbert-space representation on $\mathcal{H}_i$
\begin{align}
	&\hat{x}^\mu_{(i)} \ket{\phi_i}_X = \int d^4x \sqrt{-g_i(x)} x^\mu \phi_i(x) \ket{x^{(i)}}_X\\
	&\hat{p}_\nu^{(i)} \ket{\phi_i}_X = \int d^4x \sqrt{-g_i(x)} [-g_i(x)]^{-1/4}\frac{\partial}{\partial x^\nu}\left[ [-g_i(x)]^{1/4} \phi_i(x) \right] \ket{x^{(i)}}_X.
\end{align}
This representation on the Hilbert space guarantees that the operators have canonical commutation relations on each manifold $\mathcal{M}_i$
\begin{equation}
	\left[ \hat{x}^\mu_{(i)}, \hat{p}_\nu^{(j)} \right] = i \hbar\, \delta^\mu_\nu \delta_{ij}.
\end{equation}
This representation on the Hilbert space ensures that the operators are hermitian. 

\section{Transformation to a Locally Inertial Frame}
\label{App:LIFTrans}

We now want to describe the change to a Locally Inertial Frame (LIF) centred in a point $x_P$. Clearly, this change of coordinate is only valid locally, because it is obtained by expanding the coordinate transformation in a Taylor series about that point. For our purpose (looking at the gravitational field at the location of a pointlike quantum particle) it is enough to consider terms up to the second order
\begin{equation}
	\xi^\mu = b^\mu_\alpha (x-x_P)^\alpha + \frac{1}{2}  b^\mu_\lambda \Gamma^\lambda_{\alpha \beta} (x-x_P)^\alpha (x-x_P)^\beta + \cdots,
\end{equation}
where $b^\mu_\alpha = \frac{\partial \xi^\mu}{\partial x^\alpha}\Big |_{x_P}$ and $\Gamma^\lambda_{\alpha \beta} = \Gamma^\lambda_{\alpha \beta}(x_P)$. The inverse of this transformation is
\begin{equation}
	(x-x_P)^\alpha = f^\alpha_\mu \xi^\mu -\frac{1}{2} f^\gamma_\mu f^\beta_\nu \Gamma^\alpha_{\beta\gamma}\xi^\mu \xi^\nu + \cdots,
\end{equation}
with the $f$ matrix being the inverse of the $b$ matrix, i.e., $f= b^{-1}$. We can now apply the change of coordinates to the locally inertial frame to the metric $g_{\alpha\beta}(x)$, such that
\begin{equation}
	\tilde{g}_{\mu\nu} (\xi) = g_{\alpha\beta}(x(\xi)) \frac{\partial x^\alpha}{\partial \xi^\mu} \frac{\partial x^\beta}{\partial \xi^\nu}.
\end{equation}
Considering $\frac{\partial x^\alpha}{\partial \xi^\mu} = f^\alpha_\mu - f^\gamma_\mu f^\beta_\nu \Gamma^\alpha_{\beta\gamma} \xi^\nu + \cdots$ and by recasting the old metric in terms of the new coordinates, we find that the gauge freedom allows us to choose the entries of the matrix $f$ such that the zeroth order of the transformation is the Minkowski metric $\eta_{\mu\nu}$ \cite{Misner:1974qy}. The first order is identically zero, so overall we get
\begin{equation}
	\tilde{g}_{\mu\nu} (\xi) = \eta_{\mu\nu} + O(\xi^2).
\end{equation}

\section{Action formalism for a quantum particle in free fall in the gravitational field}
\label{App:PathIntegral}

Here, we show that we can formulate such action principle for a quantum particle in a gravitational background. We use the formalism of covariant quantum mechanics (CQM) \cite{reisenberger2002spacetime, Rovelli:2004tv} in order to treat space and time on the same footing. 

CQM is obtained from standard quantum mechanics by adding a redundancy, i.e., a coordinate time operator, which behaves as every other spatial coordinate operator. This implies that the dynamical laws of time evolution is now written as a constraint. The physical solutions are then obtained by enforcing the constraint (i.e., by passing from the Kinematical to the Physical Hilbert space). Formally, this can be written as a group-averaging operation
\begin{equation} \label{eq:Constraint}
	\ket{\psi_p} = \int d^4p \delta(\hat{C}) \psi_k (p) \ket{p^\mu} = \frac{1}{\sqrt{2\pi \hbar}}\int d\tau e^{\frac{i}{\hbar} \hat{C} \tau} \ket	{\psi_k},
\end{equation}
where $\ket{\psi_k} = \int d\mu(x) \psi_k (x) \ket{x}$ is a state in the Kinematical Hilbert space and $d\mu(x) = d^4x \sqrt{-g(x)}$, with $g(x)$ being the determinant of the metric, is the covariant integration measure. This condition is equivalent to $\hat{C}\ket{\psi_p}=0$. The operator $\hat{P} = \frac{1}{\sqrt{2\pi \hbar}}\int d\tau e^{\frac{i}{\hbar} \hat{C} \tau}$ is an improper projector, because it doesn't satisfy $\hat{P}^2 = \hat{P}$. In order to normalise the state, it is necessary to redefine the scalar product on the physical Hilbert space. Such induced scalar product $\braket{\cdot | \cdot}_{ph}$ can be shown to be (see for instance Refs.~\cite{Marolf1995Refined, Hartle:1997dc})
\begin{equation}
	\braket{\psi_p^1 | \psi_p^2}_{ph} = \braket{\phi_1 | \hat{P} | \phi_2},
\end{equation}
where $\ket{\psi_p^i} = \hat{P} \ket{\phi_i}$ for $i=1,2$ and $\braket{\cdot | \cdot}$ is the scalar product on the Kinematical Hilbert space.

In order to write Eq.~\eqref{eq:Constraint} as an action, we need to make sure that the Legendre transform of the constraint $\hat{C}$ leads, in a path-integral approach, to the classical action. This is nontrivial: the usual approach to finding a quantum constraint is to promote the classical dispersion relation to an operator-valued constraint (constraint quantisation). The main problem that one encounters in this procedure in a general spacetime is that when we promote the classical constraint $C = g_{\mu \nu}(x) p^\mu p^\mu -m^2c^2$ to an operator there are ambiguities in the order of the operators, because the coordinates in $g_{\mu \nu} (x^\alpha)$ become operators on the Hilbert space of the particle. However, a natural solution to this problem is to choose the Weyl ordering for the operators, because this ensures that, when the constraint $\hat{C}$ is inserted in the path integral, it leads to the expression equivalent to the corresponding classical Hamiltonian (and thus to the classical Lagrangian) \cite{greiner2013field}. 

We write a general state in the Physical Hilbert space as
\begin{equation}
	\ket{\psi_p} =\frac{1}{2\pi \hbar}\int d\tau \int d\mu (x_0) e^{-\frac{i}{\hbar}\hat{C}\tau} \psi_k(x_0) \ket{x_0},
\end{equation}
where $\hat{C} = \mathcal{W}( g_{\mu \nu}(x) p^\mu p^\mu -m^2c^2)$ is the Weyl-ordered operator corresponding to the classical dispersion relation. We now regularise the integral by inserting a cutoff, thus defining $T= M\Delta_m$, with $M \in \mathbb{N}$ and $\Delta_m$ very small. We can then discretise the integral and we obtain
\begin{equation}
		\ket{\psi_p} = \lim_{\substack{T,M\rightarrow \infty \\ \Delta_m \rightarrow 0}}\frac{1}{2\pi \hbar}\sum_{m=-M}^M \Delta_m \int d\mu (x_0)  e^{-\frac{i}{\hbar}\hat{C}\Delta_m m} \psi_k(x_0) \ket{x_0},
\end{equation}
where $m\Delta_m = \tau$. In order to make each time interval small, we further split every $\Delta_m$ into $N_m$ intervals, i.e., $\Delta_m = N_m \delta_{m}$. The physical state then becomes 
\begin{equation} \label{eq:PsiPath}
	\begin{split}
		\ket{\psi_p} = \lim_{\substack{T,M\rightarrow \infty \\ \Delta_m \rightarrow 0}} &\sum_{m=-M}^M \Delta_m \int d\mu (x_0) \left( e^{-\frac{i}{\hbar}\hat{C}\delta_m m} \right)^{N_m} \psi_k(x_0) \ket{x_0} \\
		= \lim_{\substack{T,M\rightarrow \infty \\ \Delta_m \rightarrow 0}} & \sum_{m=-M}^M \Delta_m \int d\mu (x_0) \cdots d\mu (x_{N_m}) \bra{x_{N_m}} e^{-\frac{i}{\hbar}\hat{C}\delta_m m} \ket{x_{N_m-1}} \cdots \\
		& \cdots \bra{x_1} e^{-\frac{i}{\hbar}\hat{C}\delta_m m} \ket{x_{0}} \psi_k(x_0) \ket{x_{N_m}}.
	\end{split}
\end{equation} 
In going from the first to the second line, we have inserted $N_m$ resolutions of the identity. Each single factor is a transition amplitude, and can be rewritten as
\begin{equation} \label{eq:SingleTransition}
	\begin{split}
		&\bra{x_n} e^{-\frac{i}{\hbar}\hat{C}\delta_m m} \ket{x_{n-1}} = \bra{x_n} \left( \mathbb{1} -\frac{i}{\hbar}\hat{C}(\hat{x}, \hat{p})\delta_m m \right) \ket{x_{n-1}} + O (\delta_m^2)= \\
	&=  \int \frac{d^4p_n}{(2\pi \hbar)^4} \braket{x_n | p_n}\bra{p_n} \left( \mathbb{1} -\frac{i}{\hbar}\hat{C}(\hat{x}, \hat{p})\delta_m m \right) \ket{x_{n-1}} + O (\delta_m^2)= \\
	&= \int \frac{d^4p_n}{(2\pi \hbar)^4} \frac{e^{\frac{i}{\hbar}p_n^\mu (x_n- x_{n-1})_\mu}}{[-g(x_n)]^{1/4}[-g(x_{n-1})]^{1/4}} \left[ 1 -\frac{i}{\hbar}\hat{C}\left(\frac{x_n + x_{n-1}}{2}, p_n\right)\delta_m m \right] + O (\delta_m^2),
	\end{split}
\end{equation}
where in going from the second to the third line we have evaluated the Hamiltonian in the mean position, ensuring that no ordering ambiguities arise\footnote{This is a standard method in Feynman path integrals, which can be found, e.g., in Ref.~\cite{greiner2013field}. In flat spacetime ordering ambiguities do not arise, but they become relevant in curved spacetime.}. Defining $\bar{x}_n = \frac{x_n + x_{n-1}}{2}$ and $\Delta x_n = x_n -x_{n-1}$ and plugging the expression for the single transition amplitude into the expression of the physical state, we obtain
\begin{equation}
	\begin{split}
		\ket{\psi_p} = &\lim_{\substack{M\rightarrow \infty \\ \Delta_m \rightarrow 0}}\frac{1}{2\pi \hbar}\sum_{m=-M}^M \Delta_m \int d\mu(x_0) d\mu(x_{N_m}) K(x_0, x_{N_m}, 0, N_m \delta_m) \psi_k(x_0) \ket{x_{N_m}},
	\end{split}
\end{equation}
where
\begin{equation}
	\begin{split}
		 & K(x_0, x_{N_m}, 0, N_m \delta_m) =  \\
		 & [-g(x_0)]^{-1/4}[-g(x_{N_m})]^{-1/4}\int d^4 x_1 \cdots d^4 x_{{N_m}-1} d^4 p_1 \cdots d^4 p_{N_m}  \Pi_{n=1}^{N_m} e^{\frac{i}{\hbar}\left( p_n \dot{x}_n - C(\bar{x}_n, p_n)\right)\delta_m m},
	\end{split}
\end{equation}
where we have used the fact that, by definition, $\lim_{\delta_m \rightarrow 0} \frac{\Delta x_n}{\delta_m m} = \lim_{\delta_m \rightarrow 0} \frac{x_n - x_{n-1}}{\delta_m m} = \dot{x}_n$ and $\delta_m m = \tau/N_m$.

We now have to take the limits $M \rightarrow \infty$, $\Delta_m \rightarrow 0$, and $N_m \rightarrow \infty,\, \forall m=1, \cdots, M$.  First of all, we notice that $\Delta_m \rightarrow 0$ also implies $\delta_m \rightarrow 0$, and that, when $N_m$ becomes large, all terms $O(\delta_m^2)$ go to zero. This means that we can use the identity
\begin{equation}
	\lim_{N\rightarrow \infty} \Pi_{n=1}^N \left( 1+ \frac{z_n}{N} \right)^N = \exp \left( \lim_{N\rightarrow \infty} \frac{1}{N}\Sigma_{n=1}^N z_n \right)
\end{equation}
to justify Eq.~\eqref{eq:SingleTransition}. We can thus write
\begin{equation}
	\begin{split}
		 & K(x_0, x(\tau), 0, \tau) = \lim_{\substack{\Delta_m \rightarrow 0 \\ N_m \rightarrow \infty}}  K(x_0, x_{N_m}, 0, \tau) =  \\
		 & = [-g(x_0)]^{-1/4}[-g(x(\tau))]^{-1/4}\int \mathcal{D}[x(\tau)]  e^{\frac{i}{\hbar} \int_0^\tau ds \mathcal{L}(x(s), \dot{x}(s), s)},
	\end{split}
\end{equation}
where $\mathcal{L}(x(s), \dot{x}(s), s) = \sqrt{g_{\mu\nu} (x(s)) \dot{x}^\mu \dot{x}^\nu} $ and the limit $N_m \rightarrow \infty$ should be intended as the limit when $N_m$ becomes large, but the product $N_m \delta_m = \Delta_m$. Finally, we define the physical state as the state obtained by sending $M \rightarrow \infty$, or equivalently $T\rightarrow \infty$, i.e.,

\begin{equation} \label{eq:actionPhys}
	\begin{split}
		\ket{\psi_p} =& \lim_{T\rightarrow \infty}\frac{1}{2\pi \hbar} \int_{-T/2}^{T/2} d\tau \int d\mu(x_0) d\mu(x(\tau))  K(x_0, x(\tau),0, \tau)\psi_k(x_0) \ket{x(\tau)}=\\
		=&\frac{1}{2\pi \hbar} \int_{\mathbb{R}} d\tau \int \mathcal{D}[x(\tau)] \exp \left\lbrace \frac{i}{\hbar} \int d\tau \mathcal{L}(x(\tau), \dot{x}(\tau), \tau) \right\rbrace  \psi_k(x_0) \ket{x(\tau)}.
	\end{split}
\end{equation}

We have thus generalised the path integral to the case in which the time evolution of the particle is encoded in a constrained (timeless) Hamiltonian in general spacetime.

\section{History state for a freely falling quantum particle in a classical gravitational field}
\label{App:HistoryState}

The physical state of a quantum particle which is freely falling in a classical gravitational field can be written, using the action formalism derived in App.~\ref{App:PathIntegral}, as
\begin{equation} \label{eq:omegatau}
	\ket{\phi_p} = \int d\tau \omega(x(\tau)) \ket{x(\tau)},
\end{equation}
where
\begin{equation}
		\omega(x(\tau)) = \frac{1}{2\pi \hbar} \int d\mu[x_0] d\mu[x(\tau)] K(x_0, x(\tau),0, \tau) \psi_k(x_0),
\end{equation}
the covariant integration measure is $d\mu(x) = \sqrt{-g(x)}d^4x$ and
\begin{equation}
	\begin{split}
		 & K(x_0, x(\tau), 0, \tau) =  [-g(x_0)]^{-1/4}[-g(x(\tau))]^{-1/4}\int \mathcal{D}[x(\tau)]  e^{\frac{i}{\hbar} \int_0^\tau ds \mathcal{L}(x(s), \dot{x}(s), s)},
	\end{split}
\end{equation}
with $\mathcal{L}(x(s), \dot{x}(s), s)$ being the classical Lagrangian  $\mathcal{L}(x(s), \dot{x}(s), s) = \sqrt{g_{\mu\nu} (x(s)) \dot{x}^\mu \dot{x}^\nu} $.

We now use the fact that the physical state is invariant by translations in $\tau$ and write
\begin{equation} \label{eq:omegaepsilon}
	\ket{\phi_p} = \int d\tau \omega(x(\tau +\epsilon)) \ket{x(\tau + \epsilon)},
\end{equation}
where $\epsilon$ is an arbitrary quantity and 
\begin{equation}
	\begin{split}
		&\omega(x(\tau + \epsilon)) = \frac{1}{2\pi \hbar} \int d\mu[x_0] d\mu[x(\tau + \epsilon)][-g(x_0)]^{-1/4}[-g(x(\tau + \epsilon))]^{-1/4}\times\\
		&\times \int \mathcal{D}[x(\tau + \epsilon)]  e^{\frac{i}{\hbar} \int_0^{\tau + \epsilon} ds \mathcal{L}(x(s), \dot{x}(s), s)} \psi_k(x_0)=\\
		&= \frac{1}{2\pi \hbar} \int d\mu[x_0]  d\mu[x(\tau)] d\mu[x(\tau + \epsilon)] K(x_0, x(\tau), 0, \tau) K(x(\tau),x(\tau + \epsilon), \tau, \tau + \epsilon)\psi_k(x_0),
	\end{split}
\end{equation}
which is equivalent to the composition rule for time evolution $\hat{U}(t, t_0) = e^{-\frac{i}{\hbar}\hat{H}(t-t_0)}$ in ordinary quantum mechanics, i.e., $\hat{U}(t_2, t_0) = \hat{U}(t_2, t_1)\hat{U}(t_1, t_0)$. We can thus write
\begin{equation}
	\begin{split}
		&\int d\mu[x(\tau + \epsilon)] K(x(\tau),x(\tau + \epsilon), \tau, \tau + \epsilon) \ket{x(\tau + \epsilon)}=\\
	=& 	\int d\mu[x(\tau + \epsilon)][-g(x(\tau))]^{-1/4}[-g(x(\tau + \epsilon))]^{-1/4} \int \mathcal{D}[x(\tau+\epsilon)] e^{\frac{i}{\hbar}\int_\tau^{\tau + \epsilon} \mathcal{L}(x(s), \dot{x}(s), s)} \ket{x(\tau + \epsilon)}.
	\end{split}
\end{equation}

We now perform a change of variables, by denoting the final time $x(\tau + \epsilon) = x'(\tau)$, and by denoting the intermediate positions for $\tau < s < \tau + \epsilon$ as $x(s) = x'(s) + \delta(s)$, where $\delta(s)$ is an arbitrary function of $s$ (notice that the extremal points are excluded, hence $\delta(\tau)=0$). We then obtain
\begin{equation}
	\begin{split}
		&\int d\mu[x(\tau + \epsilon)] K(x(\tau),x(\tau + \epsilon), \tau, \tau + \epsilon) \ket{x(\tau + \epsilon)}=\\
		=&\int d\mu[x'(\tau)] [-g(x(\tau))]^{-1/4}[-g(x'(\tau))]^{-1/4}\int \mathcal{D}[x'(\tau)] e^{\frac{i}{\hbar}\int_\tau^{\tau + \epsilon} \mathcal{L}(x'(s) + \delta(s),\, \dot{x}'(s) + \dot{\delta}(s), \,s)} \ket{x'(\tau)}=\\
		=&\int d\mu[x'(\tau)] K_\delta (x(\tau), x'(\tau), \tau, \tau+\epsilon) \ket{x'(\tau)},
	\end{split}
\end{equation}
where $K_\delta (x(\tau), x'(\tau), \tau, \tau+\epsilon)$ is
\begin{equation}
	K_\delta (x(\tau), x'(\tau), \tau, \tau+\epsilon)=[-g(x(\tau))]^{-1/4}[-g(x'(\tau))]^{-1/4}\int \mathcal{D}[x'(\tau)] e^{\frac{i}{\hbar}\int_\tau^{\tau + \epsilon} \mathcal{L}(x'(s) + \delta(s),\, \dot{x}'(s) + \dot{\delta}(s), \,s)}.
\end{equation}

Since the states of Eqs.~\eqref{eq:omegatau} and \eqref{eq:omegaepsilon} can be obtained by simply applying a translation in $\tau$ to one of them, and since the translated state is still a solution of the constraint, we have that
\begin{equation}
	\begin{split} \label{eq:zeroidentity}
		0 =& \int d\tau \left\lbrace\omega(x(\tau +\epsilon)) \ket{x(\tau + \epsilon)}-  \omega(x(\tau)) \ket{x(\tau)}\right\rbrace=\\
		 = & \frac{1}{2\pi \hbar}\int d\tau \int d\mu[x_0] d\mu[x(\tau)] K(x_0, x(\tau), 0, \tau)  \psi_k(x_0) \times \\
		 &\times \left\lbrace \int d\mu[x'(\tau)] K_\delta(x(\tau), x'(\tau), \tau, \tau + \epsilon) \ket{x'(\tau)}-  \ket{x(\tau)}\right\rbrace. 
	\end{split}
\end{equation}
By keeping the value $\epsilon$ fixed, we now perform a variation on the positions, and we write
\begin{equation} \label{eq:PerturbativeK}
	\begin{split}
		&\int d\mu[x'(\tau)] K_\delta(x(\tau), x'(\tau), \tau, \tau + \epsilon) \ket{x'(\tau)}= \int d\mu[x'(\tau)] K(x(\tau), x'(\tau), \tau, \tau + \epsilon) \ket{x'(\tau)}+\\
		& +  \int d\mu[x'(\tau)] \frac{\delta}{\delta[\delta(\tau)]} \left\lbrace K_\delta (x(\tau), x'(\tau), \tau, \tau + \epsilon)\right\rbrace_{\delta(\tau)=0}\delta(\tau) \ket{x'(\tau)}   + O(\delta(\tau)^2).
	\end{split}
\end{equation}
The first order variation is
\begin{equation}
	\frac{\delta}{\delta[\delta(\tau)]}\left\lbrace K_\delta (x(\tau), x'(\tau), \tau, \tau + \epsilon) \right\rbrace_{\delta(\tau)=0}= \frac{i}{\hbar} \frac{\delta S}{\delta[\delta(\tau)]}\Big|_{\delta(\tau)=0}K (x(\tau), x'(\tau), \tau, \tau + \epsilon)
\end{equation}
where $S =  \int_\tau^{\tau + \epsilon} ds \mathcal{L}(x'(s)+ \delta(s), \dot{x}'(s)+ \dot{\delta}(s), s) $. The evaluation of this expression involves a few subtleties. In order to avoid problems arising when the functional derivative is evaluated at the boundaries, we will evaluate it by using a limit procedure
\begin{equation}
	\frac{\delta S}{\delta[\delta(\tau)]}\Big|_{\delta(\tau)=0}  = \lim_{\delta(\tau) \rightarrow 0} \lim_{\tau_0 \rightarrow \tau } \frac{\delta S}{\delta [\delta(\tau_0)]},
\end{equation}
where $\tau < \tau_0 < \tau + \epsilon$. To start with, we calculate the variation of the action
\begin{equation}
	\begin{split}
		\delta S &= \delta \left\lbrace \int_\tau^{\tau + \epsilon} ds \mathcal{L}(x(s)+ \delta(s), \dot{x}(s)+ \dot{\delta}(s), s) \right\rbrace=\\
		&=  \int_\tau^{\tau + \epsilon} ds \left\lbrace \frac{\partial \mathcal{L}}{\partial x(s)}\delta(\delta(s))  +  \frac{\partial \mathcal{L}}{\partial \dot{x}(s)} \delta(\dot{\delta}(s)) \right\rbrace=\\
		&= \int_\tau^{\tau + \epsilon} ds \left\lbrace \frac{\partial \mathcal{L}}{\partial x(s)} -\frac{d}{ds} \frac{\partial \mathcal{L}}{\partial \dot{x}(s)} \right\rbrace \delta (\delta(s))+ \left[\frac{\partial \mathcal{L}}{\partial \dot{x}(s)} \delta (\delta(s)) \right]_\tau^{\tau + \epsilon}.
	\end{split}
\end{equation}
We can now compute
\begin{equation}
	\frac{\delta S}{\delta (\delta_{\tau_0})} = \frac{\partial \mathcal{L}}{\partial x'(\tau_0)} -\frac{d}{d \tau_0} \frac{\partial \mathcal{L}}{\partial \dot{x'}(\tau_0)},
\end{equation}
where the contribution of the border term vanishes. Hence, we find
\begin{equation}
	\frac{\delta S}{\delta[\delta(\tau)]}\Big|_{\delta(\tau)=0} = \frac{\partial \mathcal{L}}{\partial x'(\tau)} -\frac{d}{d \tau} \frac{\partial \mathcal{L}}{\partial \dot{x}'(\tau)},
\end{equation}

If we insert this expression into Eq.~\eqref{eq:zeroidentity} by making use of Eq.~\eqref{eq:PerturbativeK} we find that
\begin{equation}
	\begin{split}
		&0 = \int d\tau \int d\mu[x_0] d\mu[x(\tau)] K(x_0, x(\tau), 0, \tau)  \psi_k(x_0) \left\lbrace \int d\mu[x'(\tau)] K(x(\tau), x'(\tau), \tau, \tau + \epsilon) \ket{x'(\tau)} + \right.\\
	 & \left. +\frac{i}{\hbar} \int d\mu[x'(\tau)] K(x(\tau), x'(\tau), \tau, \tau+\epsilon) \left(\frac{\partial \mathcal{L}}{\partial x'(\tau)} -\frac{d}{d \tau} \frac{\partial \mathcal{L}}{\partial \dot{x}'(\tau)}\right) \delta(\tau) \ket{x'(\tau)}  - \ket{x(\tau)}\right\rbrace.
	\end{split}
\end{equation}

We now take the limit $\epsilon \rightarrow 0$ of the previous expression. We notice that 
\begin{equation}
	\lim_{\epsilon \rightarrow 0 } \int d\mu[x'(\tau)] K(x(\tau), x'(\tau), \tau, \tau + \epsilon) \ket{x'(\tau)} = \ket{x(\tau)}.
\end{equation}
Hence, we find
\begin{equation}
	0 = \frac{i}{\hbar} \int d\tau \int d\mu[x_0] d\mu[x(\tau)] K(x_0, x(\tau), 0, \tau)  \psi_k(x_0)\left(\frac{\partial \mathcal{L}}{\partial x(\tau)} -\frac{d}{d \tau} \frac{\partial \mathcal{L}}{\partial \dot{x}(\tau)}\right) \delta(\tau) \ket{x(\tau)}.
\end{equation}
Given that $\delta(\tau)$ is an arbitrary function, because of the fundamental lemma of calculus of variations we obtain
\begin{equation}
	0 = \int d\mu[x_0] d\mu[x(\tau)] K(x_0, x(\tau), 0, \tau)  \psi_k(x_0)\left(\frac{\partial \mathcal{L}}{\partial x(\tau)} -\frac{d}{d \tau} \frac{\partial \mathcal{L}}{\partial \dot{x}(\tau)}\right) \ket{x(\tau)}.
\end{equation}
We now use the fact that $\ket{x(\tau)}$ is a linearly independent basis of vectors in a Hilbert space, hence the expression can be zero if and only if
\begin{equation}
	0 = \int d\mu[x_0] K(x_0, x(\tau), 0, \tau)  \psi_k(x_0)\left(\frac{\partial \mathcal{L}}{\partial x(\tau)} -\frac{d}{d \tau} \frac{\partial \mathcal{L}}{\partial \dot{x}(\tau)}\right).
\end{equation}
We are now left with the product of two functions, because the expression between brackets does not depend on $x_0$. Hence, this is zero only if one of the two factors is zero. The term on the left cannot be zero in general, because it's the Green function and $\psi_k(x_0)$ is arbitrary. Hence, we obtain that this expression vanishes only if
\begin{equation}
	0 = \frac{\partial \mathcal{L}}{\partial x(\tau)} -\frac{d}{d \tau} \frac{\partial \mathcal{L}}{\partial \dot{x}(\tau)}.
\end{equation}

Overall, we find that invariance of the physical state under the propagation of the solution from $\tau$ to $\tau + \epsilon$ (which is built in the definition of the physical state) is equivalent to asking that the state $\ket{x(\tau)}$ satisfies the classical equations of motion with proper time $\tau$. This result implies that we can view the physical state 
\begin{equation}
	\ket{\phi_p} = \int d\tau \omega(x(\tau)) \ket{x(\tau)}
\end{equation}
as a linear combination in spacetime of configurations $\ket{x(\tau)}$, where at each value of $\tau$ the state $\ket{x(\tau)}$ satisfies the classical geodesic equation. Hence, the physical state can be written as a linear combination of classical geodesics.

\section{Proof of the equivalence of the history state from the point of view of the laboratory and of the quantum particle in the Newtonian limit}
\label{App:NewtonstateEquiv}

We consider the history state in the reference frame of a laboratory $L$, a quantum particle $P$ with internal degrees of freedom $I$, in a Newtonian gravitational field. We want to prove the relation of Eq.~\eqref{eq:Newton2}
\begin{equation} \label{eq:AppNewton2}
		\int dt e^{- \frac{i}{\hbar}\hat{H}_{PI} t} \ket{\psi_0}^{(L)} \ket{t}_L =  \int d\tau  e^{- \frac{i}{\hbar}\hat{H}'_{PL} \tau}\ket{\psi_0}^{(I)}\ket{\tau}_I,
\end{equation}
where
\begin{equation} \label{eq:AppHPIfromL}
	\hat{H}_{PI}= mc^2 + \frac{|\hat{\mathbf{p}}|^2}{2m} + m \Phi(\hat{\mathbf{x}}) + \hat{H}_I\left( 1 + \frac{\Phi(\hat{\mathbf{x}})}{c^2} \right),
\end{equation}
is the Hamiltonian of the systems $P$ and $I$ from the laboratory perspective, $\hat{H}_I$ is the Hamiltonian of the internal degrees of freedom $I$, $\ket{t}_L$ is the eigenstate of the time operator $\hat{T}_L$ of the laboratory clock, i.e., $\hat{T}_L \ket{t}_L = t \ket{t}_L $, with $[\hat{T}_L, \hat{H}_L ] = i\hbar$, and
\begin{equation}
	\ket{\psi_0}^{(L)}= \int dE_I d^3 x \psi_0 (\mathbf{x}, E_I) \ket{\mathbf{x}}_{P} \ket{E_I}_I.
\end{equation}
We want to show that the expression on the left-hand side of Eq.~\eqref{eq:AppNewton2} is equivalent to the expression on the right-hand side of Eq.~\eqref{eq:AppNewton2}, where 
\begin{align}
	&\hat{H}'_{PL} = \left[ mc^2 + \frac{|\hat{\mathbf{p}}|^2}{2m} +\left( 1 - \frac{\Phi(\hat{\mathbf{x}})}{c^2} \right) \hat{H}_L\right],\label{eq:AppHPI}\\
	&\ket{\psi_0}^{(I)}= \frac{1}{2\pi\hbar} \int d\tau_0 d^3 x dE_I dE_L \left( 1 - \frac{\Phi(\mathbf{x})}{c^2} \right)\psi_0 (\mathbf{x}, E_I) e^{\frac{i}{\hbar}(\hat{H}'_{PL}+ E_I )\tau_0} \ket{\mathbf{x}}_P \ket{E_L}_L \label{eq:Apppsi_0I},
\end{align} 
and $\ket{\tau}_I$ is the eigenstate of the time operator $\hat{T}_I$ of the particle's clock, i.e., $\hat{T}_I \ket{\tau}_I = \tau \ket{\tau}_I $, with $[\hat{T}_I, \hat{H}_I ] = i\hbar$.

In order to show this equivalence, we take the Fourier transform on the laboratory clock $L$ and on the particle's clock $I$, and we write
\begin{equation}
	\begin{split}
		\ket{\Psi} &= \int dt  e^{- \frac{i}{\hbar}\hat{H}_{PI} t} \ket{\psi_0}^{(L)} \ket{t}_L =\\
		&=\frac{1}{2\pi\hbar}\int dt d\tau_0 dE_I dE_L d^3 x e^{- \frac{i}{\hbar}(\hat{H}_{PI} + \hat{H}_L) t} e^{\frac{i}{\hbar}E_I \tau_0} \psi_0 (\mathbf{x}, E_I) \ket{\mathbf{x}}_{P} \ket{\tau_0}_I \ket{E_L}_L .
	\end{split}
\end{equation}
We now insert the expression of Eq.~\eqref{eq:AppHPIfromL} for $\hat{H}_{PI}$ and act on the states of clock $I$ with the clock Hamiltonian $\hat{H}_I$, which gives
\begin{equation}
	\begin{split}
		\ket{\Psi} &= \frac{1}{2\pi\hbar}\int dt d\tau_0 dE_I dE_L d^3 x e^{- \frac{i}{\hbar}(mc^2 + \frac{|\hat{\mathbf{p}}|^2}{2m} + m \Phi(\hat{\mathbf{x}}) + \hat{H}_L) t} e^{\frac{i}{\hbar}E_I \tau_0} \psi_0 (\mathbf{x}, E_I) \times\\
		&\times \ket{\mathbf{x}}_{P} \ket{\tau_0 + \left[1+ c^{-2}\Phi(\mathbf{x})\right]t}_I \ket{E_L}_L .
	\end{split}
\end{equation}
By changing variable $\tau = \tau_0 + \left[1+ c^{-2}\Phi(\mathbf{x})\right]t$, we obtain
\begin{equation}
	\begin{split}
		\ket{\Psi} &= \frac{1}{2\pi\hbar}\int d\tau d\tau_0 dE_I dE_L d^3 x \left(1-\frac{\Phi(\hat{\mathbf{x}})}{c^2}\right) e^{- \frac{i}{\hbar}\left(1-\frac{\Phi(\hat{\mathbf{x}})}{c^2}\right)\left(mc^2 + \frac{|\hat{\mathbf{p}}|^2}{2m} + m \Phi(\hat{\mathbf{x}}) + \hat{H}_L\right) (\tau-\tau_0)} e^{\frac{i}{\hbar}E_I \tau_0} \times\\
		&\times \psi_0 (\mathbf{x}, E_I)  \ket{\mathbf{x}}_{P} \ket{\tau}_I \ket{E_L}_L,
	\end{split}
\end{equation}
where, to our order of approximation, we find
\begin{equation}
	\left(1-\frac{\Phi(\hat{\mathbf{x}})}{c^2}\right)\left(mc^2 + \frac{|\hat{\mathbf{p}}|^2}{2m} + m \Phi(\hat{\mathbf{x}}) + \hat{H}_L\right) = mc^2 + \frac{|\hat{\mathbf{p}}|^2}{2m} + \left(1-\frac{\Phi(\hat{\mathbf{x}})}{c^2}\right)\hat{H}_L \equiv \hat{H}'_{PL}.
\end{equation}
We can then rewrite the whole expression as
\begin{equation}
	\begin{split}
		\ket{\Psi} &= \frac{1}{2\pi\hbar}\int d\tau d\tau_0 dE_I dE_L d^3 x \left(1-\frac{\Phi(\hat{\mathbf{x}})}{c^2}\right) e^{- \frac{i}{\hbar}\hat{H}'_{PL} (\tau-\tau_0)} e^{\frac{i}{\hbar}E_I \tau_0}\psi_0 (\mathbf{x}, E_I)  \ket{\mathbf{x}}_{P} \ket{\tau}_I \ket{E_L}_L =\\
		&= \int d\tau  e^{- \frac{i}{\hbar}\hat{H}'_{PL} \tau}\ket{\psi_0}^{(I)}\ket{\tau}_I ,
	\end{split}
\end{equation}
where $\hat{H}'_{PL}$ and $\ket{\psi_0}^{(I)}$ coincide with the expression given respectively in Eq.~\eqref{eq:AppHPI} and Eq~\eqref{eq:Apppsi_0I}. We have thus shown the equivalence of the left-hand side and the right-hand side of Eq.~\eqref{eq:AppNewton2}.

\bibliography{biblio}{}
\bibliographystyle{ieeetr}

\end{document}